\documentclass[a4paper,10pt]{article}

\usepackage{amsmath,amssymb}
\usepackage{color}
\usepackage{cite}

\newtheorem{remark}{Remark}[section]
\newtheorem{lemma}{Lemma}[section]
\newtheorem{proposition}{Proposition}[section]
\newtheorem{theorem}{Theorem}[section]
\newtheorem{corollary}{Corollary}[section]

\def\bbr{\mathbb{R}}\def\bbc{\mathbb{C}}
\def\rmi{\mathrm{i}}\def\rmd{\mathrm{d}}\def\rme{\mathrm{e}}
\def\tr{\mbox{ \rm tr\,}} 
\def\ad{\mbox{\rm ad\,}} \def\const{\mbox{\rm const\,}}
\def\mod{\mbox{\rm mod\,}}\def\Ad{\mbox{\rm Ad\,}}
\def\diag{\mbox{\rm diag\,}}\def\id{\mbox{\rm id\,}}

\def\SU{\mbox{\rm SU\,}}

\def\asl{\mbox{\rm sl\,}}

\def\su{\mbox{\rm su\,}}

\def\id{\mbox{\rm id\,}}

\def\pd#1,#2{\frac{\partial#1}{\partial#2}}
\def\v#1{\overline{#1}}
\def\d#1,#2{\frac{d#1}{d#2}}
\def\qed{\hbox{${\vcenter{\vbox{\hrule height 0.4pt\hbox{\vrule width 0.4pt height 6pt
\kern 5pt\vrule width 0.4pt}\hrule height 0.4pt}}}$}}

\begin{document}
\title{Pseudo-Hermitian Reduction of a Generalized Heisenberg Ferromagnet Equation.
I. Auxiliary System and Fundamental Properties}
\author{A. B. Yanovski$^1$ and T. I. Valchev$^2$\\
\small $^1$ Department of Mathematics \& Applied Mathematics,\\
\small University of Cape Town, Rondebosch 7700,
Cape Town, South Africa\\
\small $^2$ Institute of Mathematics and Informatics,\\
\small Bulgarian Academy of Sciences, Acad. G. Bonchev Str., 1113 Sofia, Bulgaria\\
\small E-mails: Alexandar.Ianovsky@uct.ac.za, tiv@math.bas.bg}
\date{}
\maketitle

\begin{abstract}
We consider an auxiliary spectral problem originally introduced by Gerdjikov, Mikhailov and Valchev ($\rm GMV$ system)
and its modification called pseudo-Hermitian reduction which is extensively studied here for the first time. We describe
the integrable hierarchies of both systems in a parallel way and construct recursion operators. Using the concept of 
gauge equivalence, we construct expansions over the eigenfunctions of recursion operators. This permits us to obtain the
expansions for both $\rm GMV$ systems with arbitrary constant asymptotic values of the potential functions in the
auxiliary linear problems. 
\end{abstract}

\section{Introduction}

We are going to study the auxiliary linear problem
\begin{eqnarray}\label{eq:gsys}
&&\tilde{L}\psi=(\rmi\partial_x - \lambda S)\psi=0,\qquad
S = \left(\begin{array}{ccc} 0&u&v\\ \epsilon u^*&0&0\\v^*&0&0
\end{array}\right), \qquad \lambda\in\bbc 
\end{eqnarray}
and the theory of expansions over the adjoint solutions related to it. Above $\epsilon=\pm 1$
and $*$ stands for complex conjugation. The complex valued functions $u$ and $v$ (the potentials)
are assumed to be smooth on $x\in\bbr$ and they satisfy the relation:
\begin{equation}\label{eq:uvcon}
\epsilon |u|^2+|v|^2=1.
\end{equation}
The potentials must also satisfy some asymptotic conditions when $x\to \pm\infty$ to be
discussed further in text.

The system (\ref{eq:gsys}) for $\epsilon = 1$ was introduced by Gerdjikov, Mikhailov and
Valchev \cite{GMV1} with the sign $+$ before $S$ but for reasons that will become clear
later we shall prefer this equivalent form (one just needs to change $(u,v)$ to $(-u,-v)$).

Let us adopt the following convention: if $\epsilon$ appears in formulas then it will be $+1$
or $-1$, if it is used as an index or label it will mean either $+$ or $-$. Thus we shall
denote the original Gerdjikov-Mikhailov-Valchev system by $\rm GMV_{+}$ while the general
case will be referred to as $\rm{GMV}_{\epsilon}$ system. 

According to \cite{GMV1,GMVSIGMA} the $\rm GMV_{+}$ system arises naturally when one looks for
integrable system having a Lax representation $[L,A]=0$ with $L$ and $A$ subject to
Mikhailov-type reduction requirements, see \cite{Mikh1, Mikh2, Mikh3}. This is also true
for both $\rm GMV_{\pm}$ systems. Indeed, the Mikhailov reduction group $G_0$ acting on the 
fundamental solutions of the system (\ref{eq:gsys}) could be defined as generated by the two
elements $g_1$ and $g_2$ acting in the following way:
\begin{equation}\label{eq:MG}
\begin{array}{l}
g_1(\psi)(x,\lambda)=\left[Q_{\epsilon}\psi(x,\lambda^{*})^{\dag}Q_{\epsilon}\right]^{-1},
\qquad Q_{\epsilon}=\diag(1,\epsilon,1),\\[4pt]
g_2(\psi)(x,\lambda)=H\psi(x,-\lambda)H, \qquad H=\diag(-1,1,1)
\end{array}
\end{equation}
where $\dag$ denotes Hermitian conjugation. Since $g_1g_2=g_2g_1$ and $g_1^2=g_2^2=\id$,
$G_0=\mathbb{Z}_2\times \mathbb{Z}_2$.  We shall call the reduction defined by $g_1,g_2$ for $\epsilon =1$
Hermitian while the reduction defined by $g_1,g_2$ for $\epsilon =-1$ pseudo-Hermitian. The
requirement that $G_0$ is a reduction group immediately gives that the coefficients of the operators $\tilde{L}$
and
\begin{equation}\label{eq:LApairs}
\tilde{A}=\rmi\partial_t + \sum_{k=0}^n\lambda^k \tilde{A}_k,
\qquad \tilde{A}_k\in \asl(3,\mathbb{C})
\end{equation}
 must satisfy:
\begin{equation}\label{eq:Arestr}
\begin{array}{l}
HSH= -S,\qquad H\tilde{A}_kH = (-1)^{k}\tilde{A}_k,\\[4pt]
Q_{\epsilon}S^{\dag}Q_{\epsilon}= S,\qquad
Q_{\epsilon}\tilde{A}_{k}^{\dag}Q_{\epsilon} = \tilde{A}_{k}.
\end{array}
\end{equation}

One may consider a more general form of pseudo-Hermitian reduction, i.e. one with $ Q_{\epsilon_1\epsilon_2} =
\diag(1,\epsilon_1,\epsilon_2)$, $\epsilon^2_1 = \epsilon^2_2 = 1$. However, it is easily checked this does
not give anything new compared to the pseudo-Hermitian reduction under consideration here.

As it can be checked the matrix $S$ has constant eigenvalues. We have $g^{-1}Sg=J_0$, where $g$ is  of
the form:
\begin{eqnarray}\label{eq:g-tr}
g=\frac{1}{\sqrt{2}}\left(\begin{array}{ccc} 1&0&-1\\
\epsilon u^*&\sqrt{2}v& \epsilon u^*\\v^*&-\sqrt{2}u&v^*\end{array}\right),
\qquad J_0=\diag(1,0-1).
\end{eqnarray}
In case $\epsilon=+1$ the matrix $g$ belongs to the group $SU(3)$ ($g^{\dag}=g^{-1}$) and when $\epsilon=-1$
to the group $SU(2,1)$ ($Q_{-}g^{\dag}Q_{-}=g^{-1}$). Further on we shall use the general notation $SU(\epsilon)$
referring to both cases, i.e. it implies $SU(\epsilon)\equiv SU(3)$ when $\epsilon = 1$ and
$SU(\epsilon)\equiv SU(2,1)$ when $\epsilon = -1$.

Since $g(x)\in SU(\epsilon)$, the values of $S(x)$ will be in the orbit $\mathcal{O}_{J_0}(\SU(\epsilon))$ of $J_0$
with respect to $SU(\epsilon )$ (it is a submanifold of $\rmi \mathfrak{su}(\epsilon)$). Thus
$S(x)\in \mathcal{O}_{J_0}(\SU(\epsilon))\cap \mathfrak{g}_{1}$ where $\mathfrak{g}_{1}$ is the space of the matrices
$X$ in $\asl(3,\mathbb{C})$ such that $HXH=-X$, see (\ref{eq:spltH}) for the reason for this notation. Let us also
note that conversely, if we assume that $S(x)\in \mathcal{O}_{J_0}(\SU(\epsilon))\cap \mathfrak{g}_{1}$ then on the
first place $S$ has the form as in (\ref{eq:gsys}). Next, as easily checked, the eigenvalues of the matrix $S$are
 \[\mu_1=0,\qquad\mu_2=-\mu_3=\sqrt{\epsilon|u|^2+ |v|^2}.\]
But since they coincide with $0,\pm 1$ we must have $\epsilon|u|^2+|v|^2=1$.  

Our approach to the $\rm GMV_{\pm}$ system will be based on the fact that it is gauge equivalent to a generalized
Zakharov-Shabat auxiliary systems  (GZS systems) on the algebra $\asl(3,\mathbb{C})$, see section \ref{gaugeq}. Generalized
Zakharov-Shabat systems, called Caudrey-Beals-Coifman (CBC) systems \cite{BeCo84} when $J$ is complex, are probably
the best known  auxiliary linear problems. In their most general form they are written on an arbitrary fixed simple
Lie algebra $\mathfrak{g}$ in some finite dimensional irreducible representation and have the form: 
\begin{equation}\label{eq:GZS}
L\psi=\left( {\rmi}\partial_x + q(x) -\lambda J \right) \psi = 0.
\end{equation}
Here, $q(x)$ and $J$ belong to $\mathfrak{g}$ in some fixed representation, $\psi$ belongs to the corresponding group.
The element $J$ must be such that the kernel of $\ad_J$ ($ \ad_J(X )\equiv [J,X]$, $X \in \mathfrak{g}$) is a Cartan
subalgebra $\mathfrak{h}_J$ of  $\mathfrak{g}$. The potential $q(x)$ belongs to the orthogonal complement
${\mathfrak{h}_J}^{\perp}$ of $\mathfrak{h}$ with respect to the Killing form:
\begin{equation}\label{eq:Killing}
\langle X, Y \rangle = \tr(\ad_X \ad_Y),\qquad X, Y \in \mathfrak{g}.
\end{equation}
It is assumed that $q(x)$ is sufficiently smooth  and it converges to $0$ fast enough when $x\to\pm\infty$.
The system (\ref{eq:GZS}) is called GZS (CBC) system over $\mathfrak{g}$ in canonical gauge. When the algebra is
understood from the context and as a representation is chosen the canonical one it is just called GZS (CBC) system.
The system (\ref{eq:GZS}) is gauge equivalent to
\begin{equation}\label{eq:BCGgauge}
\tilde{L}\tilde{\psi} = \left({\rmi}\partial_x - \lambda
S(x) \right) \tilde{\psi} = 0, \qquad S(x) \in {\cal O}_J.
\end{equation}
Usually, it is assumed that $\lim_{x\to \pm\infty} S(x) = J$ where the convergence is sufficiently fast. Here,
${\cal O}_J$ stands for the orbit of $J$ under the adjoint action of the group $G$ corresponding to $\mathfrak{g}$:
${\cal O}_J = \{ \tilde{X} : \tilde{X} = gJg^{-1}$, $g \in G \}$. The concept of gauge transformation, gauge
equivalent auxiliary problems and gauge equivalent soliton equations originates from the famous work \cite{ZaTakh79}
where it has been employed to solve an equation that is a classical analogue of equations describing waves in magnetic
chains (spin ${1}/{2}$). It has been shown that one of the nonlinear evolution equations (NLEEs) related to
$\tilde{L}$ is the Heisenberg ferromagnet equation
\[
S_t=\frac{1}{2 \rmi}[S,S_{xx}],\qquad S\in \rmi\su(2), \qquad S^2=\mathbf{1},
\qquad\lim_{x\to\pm\infty} S(x,t)=\diag(1,-1)
\]
being gauge equivalent to the famous nonlinear Schr\"odinger equation, see \cite{FadTakh87} for an extensive
discussion on that issue. It should be mentioned that the soliton
equations solvable through the auxiliary linear problem $\tilde{L}$ in the case $\asl(3,\mathbb{C})$ is related
a classical analog of the equation describing spin $1$ particle chains dynamics, see \cite{BoPo90}.

Later, the results of \cite{ZaTakh79} were generalized to the soliton equations hierarchies associated with $L$
and $\tilde{L}$, the conservation laws of those NLEEs, the hierarchies of their Hamiltonian structures etc. This
was achieved by generalizing the so-called AKNS approach \cite{AKNS} (generating operators or recursion operators
approach). Initially, this was done in the $\asl(2,\mathbb{C})$ case, next in the case of GZS system on arbitrary
semisimple Lie algebra \cite{GerYan85,G86,Yan87}. Now, this theory is referred to as the gauge-covariant theory of
the recursion operators related to the GZS (CBC) systems in canonical and pole gauge. For a detailed explanations
of all these issues and more references (prior to 2008), see the monograph \cite{GerViYa2008}.

For GZS system in pole gauge most of the essential issues could be reformulated from the canonical gauge. Perhaps
the main difficulty is to express explicitly all quantities depending on $q$ and its derivative through $S$ and
its derivatives. While there is a clear procedure how to achieve that goal, in each particular case the details
could be different. The procedure has been developed in detail in the PhD thesis \cite{Yan87}.  In the case of
$\asl(3,\mathbb{C}))$ in general position (with no reductions) it has been carried out in \cite{Yan93}. 

Regarding now the system $\rm GMV_{\pm}$, it was the Hermitian case, i.e. $\rm GMV_{+}$, that was mainly considered
so far. The first paper in which  $\rm GMV_{+}$ was studied in detail for the asymptotic conditions
$\lim_{x\to \pm\infty}u=0$, $\lim_{x\to\pm\infty}v=\exp{\rmi\Phi_{\pm}}$ was \cite{GMVSIGMA}. It contained discussion
of the spectral properties of the $GMV_{+}$ system, two operators whose product play the role of a recursion operator
were presented and expansions over the so-called adjoint solutions were derived. We shall also discuss these issues,
however, we want to stress on the following:
a) We shall be dealing with both $\rm GMV_{\pm}$ systems simultaneously; b) Our approach will be completely different,
based on the gauge equivalence we mentioned above. Consequently, we shall be able to consider general asymptotic
conditions -- constant limits $\lim_{x\to \pm\infty}u$ and $\lim_{x\to \pm\infty}v$; c) Our point of view on the
recursion operators when reductions are present is somewhat different from that adopted in \cite{GMVSIGMA}.

\section{Algebraic Preliminaries}

In order to proceed further, we shall need some information about the algebra $\asl(3,\mathbb{C})$ and the
involutions we are going to use which we introduce below.  All facts from the theory of the semisimple
Lie algebras we use are classical, see \cite{GoGr}. We also use the normalizations adopted in this monograph
and most of its notation.

The Lie algebra $\asl(3,\mathbb{C})$ is a simple Lie algebra of rank $2$. We shall denote its Killing form
$\tr(\ad_X\ad_Y)$ by $\langle X,Y\rangle$ where as usual $\ad_X(Y):=[X,Y]$. It is known that $\tr(\ad_X\ad_Y)={6}\tr(XY)$
which simplifies considerably the calculations. A Cartan subalgebra could be introduced using any regular
element $X\in \asl(3,\mathbb{C})$ and constructing the space $\mathfrak{h}_X=\ker\ad_X$. As it is known,
the subalgebra of the diagonal matrices is a canonical choice for Cartan subalgebra. It is also equal to
$\mathfrak{h}=\ker\ad_J$ where $J$ is any diagonal matrix $\diag(\lambda_1, \lambda_2, \lambda_3)$ with
distinct $\lambda_i$. In that case, we shall call $\mathfrak{h}$ \textit{the Cartan subalgebra}. For the
canonical choice of the Cartan subalgebra the system of roots $\Delta$ for $\asl(3,\mathbb{C})$ is
\begin{equation}\label{eq:roots}
\Delta = \{ \alpha_{i,j} = \epsilon_i - \epsilon_j, \quad i\neq j,
\quad i, j = 1,2,3 \}
\end{equation}
where $\epsilon_i$ are functionals acting on $\mathfrak{h}$ in the following way: $\epsilon_i({\rm diag}(h_1, h_2,h_3)) =
h_i $. Then the set of positive roots $\Delta_{+}$ consists of the elements $\alpha_i$:
\begin{equation}\label{eq:+roots}
\alpha_1 = \epsilon_1 - \epsilon_2, \quad 
\alpha_2 = \epsilon_2 - \epsilon_3, \quad 
\alpha_3 = \epsilon_1 - \epsilon_3 = \alpha_1 + \alpha_2 .
\end{equation}
The corresponding root vectors $E_{\alpha}$,  $\alpha\in \Delta$ together with the matrices $H_{\alpha_1}, H_{\alpha_2}$
written below:
\begin{eqnarray}\label{eq:Weil}\nonumber
&&E_{\alpha_1} = \frac{1}{\sqrt{6}} e_{12},\quad E_{\alpha_2} =  \frac{1}{\sqrt{6}} e_{23},
\qquad E_{\alpha_3} = \frac{1}{\sqrt{6}} e_{13},\\
&&E_{-\alpha_1} = \frac{1}{\sqrt{6}} e_{21},\qquad
E_{-\alpha_2} = \frac{1}{\sqrt{6}} e_{32},\qquad
E_{-\alpha_3} = \frac{1}{\sqrt{6}}e_{31},\\ \nonumber
&&H_{\alpha_1} = \frac{1}{6}(e_{11} - e_{22}), \qquad H_{\alpha_2} = \frac{1}{6}(e_{22} - e_{33})
\end{eqnarray}
form the Cartan-Weil basis of $\asl(3,\mathbb{C})$ associated with the Cartan subalgebra $\mathfrak{h}$.
Here as usual $e_{ij}$ means a matrix whose only nonzero entry equal to one is located in the intersection of
the $i$-th row and $j$-th column. The matrices $H_{\alpha_1}, H_{\alpha_2}$ span $\mathfrak{h}$ and the
matrices $E_{\alpha}$ span $\mathfrak{h}^{\perp}$.

The relations (\ref{eq:Arestr}) have the following Lie algebraic meaning. First, the map $h:X\mapsto HXH=HXH^{-1}$
is involutive automorphism of $\asl(3,\mathbb{C})$. Next, $Q_{\epsilon}$ defines a complex conjugation
$\sigma_{\epsilon}$  of $\asl(3,\mathbb{C})$: 
\begin{equation}
\sigma_{\epsilon}(X)=-Q_{\epsilon}X^{\dag}Q_{\epsilon}.
\end{equation}
The complex conjugation $\sigma_{\epsilon}$ defines the real form $\su(3)$ ($\epsilon=+1$) or the real
form $\su(2,1)$ ($\epsilon=-1$) of  $\asl(3,\mathbb{C})$. In order to treat both cases simultaneously we
shall adopt the notation $\su(\epsilon)$ meaning $\su(3)$ when $\epsilon=+1$ and  $\su(2,1)$ when $\epsilon=-1$.
Note that the complex conjugation $\sigma_{\epsilon}$ commutes with the automorphism $h$.  Now, let us introduce
the spaces:
\begin{equation}\label{eq:INsp}
\mathfrak{g}_j=\{X: h(X)=(-1)^{j}X\}, \quad j=0,1.
\end{equation}
Then we shall have the splittings
\begin{equation}\label{eq:spltH}
\begin{array}{l}
\asl(3,\mathbb{C})= \mathfrak{g}_{0}\oplus \mathfrak{g}_{1}\\[4pt]
\su(\epsilon)=(\mathfrak{g}_{0}\cap\su(\epsilon))\oplus ( \mathfrak{g}_{1}\cap\su(\epsilon)).
\end{array}
\end{equation}
The space $\mathfrak{g}_1$ consists of all off-diagonal matrices $X=(x_{ij})$ for which $x_{23}=x_{32}=0$ and
$\mathfrak{g}_0$ consists of all traceless matrices $X=(x_{ij})$ for which $x_{12}=x_{21}=x_{13}=x_{31}=0$.
Of course, $S$ involved in the $\rm GMV_{\pm}$ satisfies $S\in \mathfrak{g}_1$.  Also, since $h$ is automorphism,
the spaces $\mathfrak{g}_0$ and $\mathfrak{g}_1$ are orthogonal with respect to the Killing form.

Below we shall see that $h$ we already introduced ($X\mapsto h(X)=HXH$) is closely related to the automorphism
 $X\mapsto k(X)=KXK$ where 
\begin{eqnarray}\label{eq:K}
K=\left(\begin{array}{ccc} 0&0&1\\ 0&1& 0\\1&0&0\end{array}\right).
\end{eqnarray}
So we shall need how $h$ and $k$ act on the Cartan-Weil basis. Note that similar to $h$ the automorphism $k$
is also involutive, that is $k^2=\id$.  For $h$  we get that
\begin{equation}
\begin{array}{l}
h(E_{\pm\alpha_1}) = - E_{\pm \alpha_1},\qquad
h(E_{\pm\alpha_2}) =  E_{\pm \alpha_2},\qquad
h(E_{\pm\alpha_3}) = - E_{\pm \alpha_3},\\[4pt]
h(H_1)=H_1,\qquad h(H_2)=H_2.
\end{array}
\end{equation}
while for $k$ we have
\begin{equation} 
\begin{array}{l}
k(E_{\pm \alpha_{1}})=E_{\mp\alpha_2}, \qquad k(E_{\pm \alpha_{3}})=E_{\mp\alpha_3},
\qquad k(E_{\pm \alpha_{2}})=E_{\mp \alpha_1},\\[4pt]
k(H_{1})=-H_2,\qquad k(H_{2})=-H_1.
\end{array}
\end{equation}
The above formulas become simpler if we introduce action ${\cal K}$ of $k$ on the roots: 
\begin{eqnarray}
&&{\cal K}(\pm \alpha_{1})={\mp\alpha_2}, \qquad {\cal K}(\pm \alpha_{3})=\mp\alpha_3,
\qquad {\cal K}(\pm \alpha_{2})=\mp \alpha_1.
\end{eqnarray}
We observe that ${\cal K}$ maps positive roots into negative and vice-versa. Of course,  ${\cal K}$ just
as $k$ satisfies ${\cal K}^2=\id$. There is no need to introduce action of $h$ on the roots since this is
simply the identity. With the above notation we have
\begin{equation}
k(E_{\alpha})=E_{{\cal K}\alpha},\qquad h(E_{\alpha})=r(\alpha)E_{\alpha}
\end{equation}
where $r(\alpha)=1$ if $\alpha=\pm\alpha_1,\pm\alpha_3$ and $r(\alpha)=-1$ if $\alpha=\pm\alpha_2$.
Note that $r(\alpha)r(-\alpha)=1$.

The invariance under the group generated by $g_1,g_2$ means that if $\psi$ is the common $G_0$-invariant
fundamental solution of the linear problem 
\begin{equation}\label{eq:GMVgauge}
\tilde{L}\tilde{\psi} = \left(\rmi\partial_x - \lambda
S(x) \right) \tilde{\psi} = 0, \qquad S(x) \in {\cal O}_{J_0}\subset \asl(3,\mathbb{C})
\end{equation}
and the linear problem of the type:
\begin{equation}
\tilde{A}\tilde{\psi}= \rmi\partial_t\tilde{\psi} + \sum_{k=0}^n\lambda^k \tilde{A}_k\tilde{\psi}=0,
\qquad \tilde{A}_k\in \asl(3,\mathbb{C})
\end{equation}
we must have $S\in \mathfrak{g}_1\cap \rm i \su(\epsilon)$ and
\begin{eqnarray}
\tilde{A}_{2k+1}\in \mathfrak {g}_1\cap \rm i\su(\epsilon),\qquad 
\tilde{A}_{2k}\in \mathfrak {g}_0\cap \rm i\su(\epsilon), \qquad k = 0,1,2,\dots
\end{eqnarray}
The conditions on $\tilde{A}_{k}$ and $S$ coincide with (\ref{eq:Arestr}), in particular, we have that  $h(S)=-S$
and $\sigma_{\epsilon}(S)=-S$. This forces $S$ to be in the form we introduced in (\ref{eq:gsys}).  Also, as a direct
consequence from the last relations we obtain that
\begin{equation}\label{eq:f-spaces}
h\circ\ad_S=-\ad_S\circ h,\qquad \sigma_{\epsilon}\circ\ad_S=-\ad_S\circ \sigma_{\epsilon}.
\end{equation}
Consequently, the spaces $\ker\ad_S={\mathfrak{h}_S}$ (which obviously is a Cartan subalgebra) and its orthogonal
complement $\mathfrak{h}_S^{\perp}$ are invariant under $h$. Each of them splits into two $h$-eigenspaces (for
the eigenvalues $+1$ and $-1$ respectively):
\begin{eqnarray}
&&\mathfrak{h}_S^{\perp}=\mathfrak{f}_0\oplus \mathfrak{f}_1,\qquad \mathfrak{h}_S=\mathfrak{h}_0\oplus \mathfrak{h}_1.
\end{eqnarray}
Several consequences follow from the above :
\begin{enumerate}
\item $\ad_S$ and $\ad_S^{-1}$ interchange the spaces $\mathfrak{f}_0$ and $\mathfrak{f}_1$:
\begin{equation}
\begin{array}{l}
\ad_S\mathfrak{f}_0= {\mathfrak{f}}_1,\qquad \ad_S{\mathfrak{f}}_1= {\mathfrak{f}}_0,\\[4pt]
\ad_S^{-1}{\mathfrak{f}}_0= {\mathfrak{f}}_1,\qquad \ad_S^{-1}{\mathfrak{f}}_1={\mathfrak{f}}_0.
\end{array}
\end{equation}
\item Since $S\in \mathfrak{g}_1$, $S_1=S^2-\frac{2}{3}\mathbf{1}\in \mathfrak{g}_0$, the spaces
${\mathfrak{h}}_0$, ${\mathfrak{h}}_1$ are $1$-dimensional and are spanned by $S$ and
$S_1=S^2-\frac{2}{3}\mathbf{1}$ respectively.
\item From the previous item follows that
\begin{equation}
S_x\in \mathfrak{g}_1, \qquad (S_1)_x=(S^2)_x\in \mathfrak{g}_0
\end{equation}
and since $S_x$ and $(S_1)_x$ are orthogonal to ${\mathfrak{h}_S}$ we have
\begin{equation}
S_x\in {\mathfrak{f}}_1,\qquad S_{1x}\equiv (S_1)_x=(S^2)_x\in {\mathfrak{f}}_0.
\end{equation}
\end{enumerate}

Another issue we must discuss is the relation between $h$ from one side and $\ad_S^{-1}$ and $\pi_S$
from the other. Here $\pi_S$ is the orthogonal projection on the space $\mathfrak{h}_S^{\perp}$ with
respect to the Killing form and $\ad_S^{-1}$ is defined only on the space $\mathfrak{h}_S^{\perp}$.
As one can show, see \cite{GMVSIGMA,YanJMP2011},  $\ad_S^{-1}$ could be expressed as a polynomial
$p_3(\ad_S)$ in $\ad_S$ where only odd degrees of $\ad_S$ are present ($\ad_{L_1}^{-1}$ through $\ad_{L_1}$
in the notation of \cite{GMVSIGMA}). From the other side, this means that $\pi_S$  could be expressed
as a polynomial in $\ad_S$ where enter only the even powers of $\ad_S$. Consequently, one obtains that
\begin{equation}\label{eq:symSh}
\ad_S^{-1}\circ h=-h\circ \ad_S^{-1},\quad \pi_S\circ h=-h\circ \pi_S
\end{equation}
and in the same way
\begin{equation}\label{eq:SymSs}
\ad_S^{-1}\circ  \sigma_{\epsilon} =- \sigma_{\epsilon}\circ \ad_S^{-1},\qquad
\pi_S\circ  \sigma_{\epsilon}=- \sigma_{\epsilon}\circ \pi_S.
\end{equation}
\begin{remark}\label{Rem:adS}
If we understand $\ad_S^{-1}$ as polynomial $p_3(\ad_S)$ it is clear that instead of $\ad_S^{-1}\circ \pi_S$
one can simply write $\ad_S^{-1}$ which is usually done without mentioning. 
\end{remark}
Taking into account that  $\pi_S$ could be written as $p_3(\ad_S)\ad_S$, that is, as a polynomial on $\ad_S$
in which only the even degree terms are present, one sees that $\pi_S$ commutes not only with $h$ (which we
have seen already) but also with $\sigma_{\epsilon}$. For convenience we formulate these facts as a proposition:
\begin{proposition}\label{prop:commpisigh}
The projection $\pi_S$ commutes with $h$ and $\sigma_{\epsilon}$.
\end{proposition}

\section{The $\rm GMV_{\pm}$ System and its Gauge Equivalent}\label{gaugeq}

It has been pointed out \cite{GMVSIGMA} that the $\rm GMV_{+}$ system is gauge equivalent to a GZS
type system. For the system  $\rm GMV_{\pm}$ the situation is the same. After the gauge transformation
$\tilde{L}\mapsto L'=g^{-1}(x)\tilde{L}g(x)$ with $g(x)$ given in  (\ref{eq:g-tr}) we obtain the system
\begin{eqnarray}
L'\psi'=(\rmi\partial_x+q'-\lambda J_0){\psi}'=0
\end{eqnarray}
where $q'=\rmi g^{-1}g_x$.  After expressing $g$ with $u$ and $v$,
\begin{eqnarray}\label{eq:qpr}
&&q'=
\frac{\rmi}{2} \left(\begin{array}{ccc} \epsilon uu_x^*+vv_x^*&  \sqrt{2} (uv_x-vu_x)&\epsilon uu_x^*+vv_x^*\\
\sqrt{2}\epsilon(v^*u_x^*-u^*v_x^*)& 2(\epsilon u^*u_x+v^*v_x)&\sqrt{2}\epsilon(v^*u_x^*-u^*v_x^*)\\
\epsilon uu_x^*+vv_x^*&\sqrt{2}(uv_x-vu_x)&\epsilon uu_x^*+vv^*_x
 \end{array}\right).
\end{eqnarray}
As it should be for GZS system $J_0$ is a real diagonal matrix and its diagonal elements are ordered to
decrease going down along the main diagonal. This is in fact was our motivation when we choose the sign
in front of $S$. As one could see, in the above $q'$ is not off-diagonal as GZS requires and in \cite{GMVSIGMA}
authors just stopped here. It turns out, however, that the issue could be easily addressed making one
more gauge transformation. Indeed, due to the condition $\epsilon|u|^2+|v|^2=1$ the diagonal part of
$q'$ equals $\frac\rmi{2}(\epsilon uu_x^*+vv_x^*)J'=b(x)J'$ where $J'=\diag (1,-2,1)$. Moreover, the
diagonal part is real, that is $b(x)=\frac\rmi{2}(\epsilon uu_x^*+vv_x^*)$ is real. Then the gauge transformation
\begin{equation}
L'\mapsto L = \left(\exp{(-\rmi J'\int_{-\infty}^x b(y){\rmd}y)}\right) L' \left(\exp{(\rmi J'\int_{-\infty}^x b(y){\rmd}y)}\right)
\end{equation}
maps the spectral problem into
\begin{eqnarray}\label{eq:GZSsl3}
L\psi=(\rmi\partial_x+q-\lambda J_0){\psi}=0
\end{eqnarray}
where
\[q(x) = \exp\left[-\rmi J'\int_{-\infty}^x b(y)\rmd y\right]q'(x)\exp\left[\rmi J'\int_{-\infty}^x b(y)\rmd y\right].\]
One can easily see that $q$ is off-diagonal and since the entries of $q'$ decay when $x\to\pm\infty$ the entries of
$q$ also decay.  Now, since the diagonal elements of $J_0$ are distinct we conclude that $\rm GMV_{\pm}$ is a problem
of GZS type in pole gauge on $\asl(3,\mathbb{C})$. We shall exploit this fact but we want to address first some other
issues. The first one is about whether we can be more precise about the potential $q$ in the GZS system (\ref{eq:GZSsl3})
to which  $\rm GMV_{\pm}$ is equivalent.  With direct calculation we are able to establish that the potential $q$ in
(\ref{eq:GZSsl3}) has special properties, namely $Q_{\epsilon}q^{\dag}Q_{\epsilon}=q$ and $KqK=q$ where $K$ has been
introduced in (\ref{eq:K}). We also notice that $KJ_0K=-J_0$ and $Q_{\epsilon}J_0^{\dag}Q_{\epsilon}=J_0$. 

The next issue is about the asymptotic conditions $S$ must satisfy. Passing from GZS system in canonical gauge
(\ref{eq:GZSsl3}) to GZS in pole gauge (\ref{eq:BCGgauge}) usually involves the Jost solution $\psi_0'$ of
(\ref{eq:GZSsl3}) such that $\lim_{x\to-\infty}\psi_0'(x)=\mathbf{1}$ at $\lambda=0$. The gauge transformation
is then $\psi\to \tilde{\psi}=\psi_0'^{-1}\psi$. In that case we obtain $S=\psi_0'^{-1}J_0\psi_0'$ and
$\lim_{x\to-\infty}S(x)=J_0$. However, this is not our case and such a condition is not compatible with the
reductions.  Natural for the ${\rm GMV_{\pm}}$ system are conditions when $u,v$ tend to some constant values.
In this work we shall assume that $\lim_{x\to\pm\infty}u(x)=u_{\pm}$, $\lim_{x\to\pm\infty}v(x)=v_{\pm}$. In
that case the limit of $g$ when $x\to\pm\infty$ is equal to
\begin{eqnarray}\label{eq:gtrminf}
g_{\pm}=\frac{1}{\sqrt{2}}\left(\begin{array}{ccc} 1&0&-1\\ \epsilon u_{\pm}^*&\sqrt{2}v_{\pm}&  \epsilon u_{\pm}^*\\ v_{\pm}^*&-\sqrt{2}u_{\pm}&v_{\pm}^*\end{array}\right).
\end{eqnarray}
Now we notice the following. If we come back to how $q$ was constructed, we shall see that the function
\begin{equation}\label{eq:psi0}
\psi_0 = \rme^{-\displaystyle\rmi J'\int_{-\infty}^xb(y){\rmd}y}g^{-1}
\end{equation}
(for the definitions of $g$ and $b$ see the explanations after (\ref{eq:qpr})) is as a matter of fact a solution
to the system (\ref{eq:GZSsl3}) for $\lambda = 0$. It satisfies:
\begin{equation}\label{eq:psi0inf}
\lim_{x\to-\infty}\psi_0=g_{-}^{-1},\qquad \lim_{x\to+\infty}\psi_0=\exp{(-\rm i\Phi J')} g_{+}^{-1}
\end{equation}
where
\begin{equation}
\Phi=\int_{-\infty}^{+\infty}b(y)\rmd y \qquad (\mod{2\pi}).
\end{equation}
\begin{remark}\label{Rem:intmot}
One can easily show that $b(x)$ is a conservation law density for all the NLEEs associated to the $\rm GMV_{\pm}$
system. Thus $\Phi$ is a conservation law and does not depend on time. Moreover, as it is well-known the
conservation laws densities are local, this shows
\[B(x,t)=\int_{-\infty}^{x}b(y,t)\rmd y \]
is a local expression (depends on $x,t$ only through $u,v$ and its derivatives).
\end{remark}

So we can make a gauge transformation where instead of $\psi_0'$ we use $\psi_0$.  Since $J_0$ and $J'$
commute, we  obtain that $S(x)=\psi_0^{-1}J_0\psi_0$ satisfies 
\begin{equation}
\lim_{x\to -\infty}S(x)=g_{-}J_0g_{-}^{-1}=S_{-}, \qquad \lim_{x\to +\infty}S(x)=g_{+}J_0g_{+}^{-1}=S_{+}
\end{equation}
\begin{eqnarray}\label{eq:Sinf}
S_{\pm}=\left(\begin{array}{ccc} 0&u_{\pm}&v_{\pm}\\  \epsilon u_{\pm}^*&0& 0\\ v_{\pm}^*&0&0\end{array}\right),
\end{eqnarray}
i.e. we have the correct asymptotic for $S(x)$. The last remark, which is in fact quite important, is that 
taking the explicit form of $\psi_0$, see (\ref{eq:psi0}), one gets the following important formula:
\begin{equation}\label{eq:sympsio}
K\psi_0=\psi_0H, \qquad \psi_0^{-1}K=H\psi_0^{-1}
\end{equation}
where  $H$ was introduced earlier in relation to the $\rm GMV_{\pm}$ system and $K$ is given in (\ref{eq:K}).
One could also obtain the same result in another way. Suppose one starts with the GZS system and assume it
has the property that if $\psi_0$ is a fundamental solution, then  $K\psi_0K$ also is. Then naturally
$K\psi_0$ is also a fundamental solution. So we must have $K\psi_0=\psi_0P$ where $P$ is some non-degenerate
matrix. Taking $x\to -\infty$ we get that $Kg_{-}^{-1}=g_{-}^{-1}P$. But, since $Kg_{-}^{-1}=g_{-}^{-1}H$ we
have $P=H$. Finally, we  formulate the results of the above discussions as a theorem:

\begin{theorem}\label{th:GEq}
The $\rm GMV_{\pm}$ system is gauge equivalent to a canonical {\rm GZS} linear problem on $\asl(3,\mathbb{C})$
\begin{equation}\label{eq:GZSours}
L\psi=(\rmi\partial_x +q-\lambda J_0)\psi=0 
\end{equation}
subject to a Mikhailov reduction group  generated by the two elements $h_1$ and $h_2$. These elements act
on the fundamental solutions  $\psi$ of the system (\ref{eq:GZSours}) in the following way:
\begin{equation}\label{eq:MGnew}
\begin{array}{l}
h_1(\psi)(x,\lambda)=\left[Q_{\epsilon}\psi(x,\lambda^{*})^{\dag}Q_{\epsilon}\right]^{-1},
\qquad Q_{\epsilon}=\diag(1,\epsilon,1),\quad \epsilon=\pm 1,\\[4pt]
h_2(\psi)(x,\lambda)=K\psi(x,-\lambda)K.
\end{array}
\end{equation}
Since $h_1^2=h_2^2=\id$ and $h_1h_2=h_2h_1$ we have again a $\mathbb{Z}_2\times \mathbb{Z}_2$ reduction.
\end{theorem}
The properties of the fundamental solutions of the auxiliary systems play a paramount role in the spectral
theory of such systems and for the theory of integration of the NLEEs associated with them. Since for the
GZS systems with $\mathbb{Z}_p$ reductions these questions have been studied in detail, see \cite{GYaSAM2015}
(whether or not $K$ is Coxeter  automorphism does not make any difference here) the above theorem
opens the possibility to study the $\rm GMV_{\pm}$ system using its gauge-equivalent GZS system with
$\mathbb{Z}_2\times \mathbb{Z}_2$ reduction. In  \cite{GYaSAM2015}, it was considered much more complicated
case of a CBC system and the automorphism has order $p>1$ (in our case $p=2$ since $K^2=\mathbf{1}$). Then
the complex plane is divided into $p$ sectors by straight lines through the origin and in each sector there
is a fundamental analytic solution to the corresponding CBC system. In our case the things are much simpler:
we have the real line dividing $\mathbb{C}$ into upper and lower half-plane. Then the system (\ref{eq:GZSours})
possesses fundamental analytic solutions (FAS) $\chi^{\pm}(x,\lambda)$ of the form
$m^{\pm}(x,\lambda) \exp{(-\rmi\lambda xJ_0)}$. The functions $m^{+}(x,\lambda)$ ($m^{-}(x,\lambda)$) are meromorphic
in $\lambda$ in the upper (lower) half plane $\mathbb{C}_+$ ($\mathbb{C}_-$). Naturally, $\chi^{\pm}(x,\lambda)$
have the same analytic properties in $\lambda$ as $m^{\pm}(x,\lambda)$.  The points of $\mathbb{C}$ that belong
to the discrete spectrum are those at which $m^{\pm}(x,\lambda)$ and their inverse have singularities with respect
to $\lambda$ in $\mathbb{C}_+$ ($\mathbb{C}_-$). It is usually considered the case when these singularities are
poles, see for example \cite{GYa94} in the general case and \cite{GYaSAM2015} for the case of reductions. There
is no difficulty to include the discrete spectrum for the $\rm GMV_{\pm}$  in our considerations but this will
make all the formulas much more complicated so we shall do it elsewhere. Here we intend to explain mainly how our
approach works and most of the issues are algebraic. Thus, in what follows we shall assume that there is no discrete
spectrum for the GZS system gauge equivalent $\rm GMV_{\pm}$ and consequently no discrete spectrum for $\rm GMV_{\pm}$.

Continuing with the properties of $m^{\pm}(x,\lambda)$, if the potential $q(x)$ has integrable derivatives up to
the $n$-th order then
\begin{equation}\label{eq:asumplam}
m^{\pm}(x,\lambda)={\mathbf 1}+\sum\limits_{i=1}^{n}a_i(x) {\lambda}^{-i}+
{\rm o}({\lambda}^{-n})
\end{equation}
when $|\lambda|\to\infty$ and $\lambda$ remains in $\mathbb{C}_{+}$ ($\mathbb{C}_{-}$ respectively). The asymptotic
is uniform in $x\in \mathbb{R}$, and the coefficients $a_i(x)$ could be calculated through $q$ and its x-derivatives.
In particular, for absolutely integrable $q$ we have $ \lim\limits_{|\lambda| \to \infty}m^{\pm}(x,\lambda)={\mathbf 1}$.
The functions  $m^{\pm}$ allow extensions by continuity to the real line, these extensions will be denoted by the
same letters. In addition, $m^{\pm}$ satisfy:
\begin{equation}
\begin{array}{l}
Km^{\pm}(x,\lambda)K=m^{\mp}(x,-\lambda),\\[4pt]
\lim_{x\to-\infty}m^{\pm}(x,\lambda)=\mathbf{1}.
\end{array}
\end{equation}
Since $KJ_0K=-J_0$, the solutions  $\chi^{\pm}$ also allow extensions by continuity to the real line
(denoted again by the same letters) and satisfy:
\begin{eqnarray}
&&K\chi^{\pm}(x,\lambda)K=\chi^{\mp}(x,-\lambda).
\end{eqnarray}
It is simple to see that in our case $\chi^{\pm}$ will also satisfy:
\begin{equation}
Q_{\epsilon}(\chi^{\pm})^{\dag}(x,\lambda)Q_{\epsilon}=(\chi^{\mp}(x,\lambda^*))^{-1}.
\end{equation}
We build from $\chi^{\pm}(x,\lambda)$ fundamental analytic solutions $\tilde{\chi}^{\pm}(x,\lambda)$ of
the system (\ref{eq:GMVgauge}) by setting:
\begin{equation}\label{eq:FAStilde}
\tilde{\chi}^{\pm}(x,\lambda)=\psi_0^{-1}\chi^{\pm}(x,\lambda).
\end{equation}
Now we have:

\begin{theorem}\label{fas_sym}
The solutions $\tilde{\chi}^{\pm}(x,\lambda)$ satisfy\footnote{In \cite{GMV1,GMVSIGMA} FAS have mistakenly been claimed to satisfy $H\tilde{\chi}^{\pm}(x,\lambda)H = \tilde{\chi}^{\mp}(x,-\lambda)$.}:
\begin{equation}
Q_{\epsilon}(\tilde{\chi}^{\pm}(x,\lambda^*))^{\dag}Q_{\epsilon}=(\tilde{\chi}^{\mp}(x,\lambda))^{-1},
\qquad H\tilde{\chi}^{\pm}(x,\lambda)H = \tilde{\chi}^{\mp}(x,-\lambda)KH.
\end{equation}
\end{theorem}
{\it Proof}. The proof of the first statement is straightforward, one needs to see only that
$Q_{\epsilon}J_0^{\dag}Q_{\epsilon}=J_0$. The second one is obtained if one takes into account (\ref{eq:sympsio}):
\begin{eqnarray*}
H\tilde{\chi}^{\pm}(x,\lambda)=H\psi_0^{-1}\chi^{\pm}(x,\lambda)=\psi_0^{-1}K\chi^{\pm}(x,\lambda)\\
= \psi_0^{-1}K\chi^{\pm}(x,\lambda)K^2=\psi_0^{-1}\chi^{\mp}(x,-\lambda)K=\tilde{\chi}^{\pm}(x,-\lambda)K.
\end{eqnarray*}
Finally, we note that we have the same type of asymptotic behavior both for $\rm GMV_{+}$ and $\rm GMV_{-}$
which allowed to treat both cases simultaneously. Of course, the solutions we speak of are different for
different choices of $\epsilon$ but we denote them by the same letter since it will not cause ambiguities.
As it is well known, the asymptotic of $\chi^{\pm}(x,\lambda)$ when $x\to\pm\infty$ are of the
type  $a_{\pm}^{\pm}\exp{(-\rmi\lambda J_0x)}$. So the asymptotic of $\tilde{\chi}^{\pm}(x,\lambda)$ are of
the type $g_{-}a^{\pm}_{-}\exp{(-\rmi\lambda J_0x)}$ when $x\to -\infty$ and
$g_{+}\exp{\rmi(\Phi J')}a^{\pm}_{+}\exp{-\rmi{\lambda J_0x}}$ when $x\to +\infty$.

\section{Recursion Operators for the $\rm GMV_{\pm}$ System}

The recursion operators ($\Lambda$-operators) are theoretical tools that permits to describe \cite{GerViYa2008}:
\begin{itemize}
\item The hierarchies of the NLEEs related to the auxiliary linear problem. 
\item The hierarchies of conservation laws for these NLEEs.
\item The hierarchies of compatible Hamiltonian structures of these NLEEs.
 \end{itemize}
The issue about the Hamiltonian structures is closely related to a beautiful interpretation of the
recursion operators originating from the work \cite{Mag78}. In fact, the adjoint of the recursion operators
could be interpreted as Nijenhuis tensors on the infinite dimensional manifold of `potentials'
(the functions $S(x)$), see \cite{GerViYa2008}.

Recursion operators arise naturally when one tries to find the hierarchy of Lax pairs related to a
particular auxiliary linear problem \cite{ZaKo85}. So suppose  we want to find the NLEEs having Lax
representation $[\tilde{L},\tilde{A}]=0$ with $\tilde{L}=\rmi\partial_x-\lambda S$ as in (\ref{eq:GMVgauge})
and 
\begin{eqnarray}\label{eq:pLax}
\tilde{A}=\rmi\partial_t+\sum\limits_{k=0}^n\lambda^k\tilde{A}_k, \qquad \tilde{A}_n\in \mathfrak{h}_S.
\end{eqnarray}
The first thing we notice is that we must have $\rmi\partial_x \tilde{A}_n \in \mathfrak{h}_S^{\perp}$ and $\tilde{A}_0=\const$
in $x$. Using a gauge transformation depending only on $t$ one can achieve $\tilde{A}_0=0$.  Then one can
show that the coefficients $\tilde{A}_k$ for $k=1,2,\ldots n-1$ are calculated recursively.  In order to
see it we first recall that the elements $S$ and $S_1 = S^2-\frac{2}{3}\mathbf{1}$
span $\mathfrak{h}_S=\ker\ad_S$ and that
\begin{equation}
\langle S,S\rangle=12, \qquad  \langle S_1,S_1\rangle=4,\qquad \langle S, S_1\rangle=0.
\end{equation}
As before, denote by $\pi_S$ the orthogonal projection (with respect to the Killing form) onto the space
$\mathfrak{h}^{\perp}_S$. Then the orthogonal projection onto $\mathfrak{h}_S$ will be $\mathbf{1}-\pi_S$
and for $X\in \asl(3,\mathbb{C})$ we shall have 
\begin{equation}
(\mathbf{1}-\pi_S)X=\frac{S}{12}\langle S,X\rangle+\frac{S_1}{4}\langle S_1,X\rangle .
\end{equation}
Putting
\begin{equation} 
\tilde{A}_s^{\rm{d}}=(\mathbf{1}-{\pi}_S)\tilde{A}_s, \quad \tilde{A}_s^{\rm{a}}=\pi_S\tilde{A}_s,
\end{equation}
we obtain  that the coefficients $\tilde{A}_k$ for $k=1,2,\ldots n-1$ are obtained recursively in the following way:
\begin{eqnarray}\label{eq:d1}
&&\tilde{A}_{k-1}^{\rm{a}}=\tilde{\Lambda}_{\pm}\tilde{A}_{k}^{\rm{a}},\\ \label{eq:d2}
&&\tilde{A}_{k}^{\rm{d}}=\rmi\displaystyle\left\{ \frac{S}{12}\int\limits_{\pm\infty}^x \langle \tilde{A}_{k}^{\rm{a}},
S_y \rangle \rmd y + \frac{S_1}{4}\int\limits_{\pm\infty}^x
\langle \tilde{A}_{k}^{\rm{a}}, S_{1y} \rangle \rmd y \right\}\\ \nonumber
\end{eqnarray}
where
\begin{eqnarray}\label{eq:ROGMV}
&&\tilde{\Lambda}_\pm(\tilde{Z})= \rmi {\ad}_S^{-1} {\pi}_S
\displaystyle\left( \partial_x\tilde{Z} + \frac{S_x}{12}\int\limits_{\pm\infty}^x \langle \tilde{Z}, S_y \rangle \rmd y + \frac{S_{1x}}{4}\int\limits_{\pm\infty}^x
\langle \tilde{Z}, S_{1y} \rangle \rmd y \right)
\end{eqnarray}
and for the sake of brevity we write here and below $S_{1x}$ instead of $(S_1)_x$.

The operators $\tilde{\Lambda}_\pm$ are recursion (generating) operators for $\rm GMV_{\pm}$
system. As one can see, they do not depend on the second reduction, that is on the choice of the real
form. In fact, the recursion operators for the $\asl(3,\mathbb{C})$-GZS system in general position
were already known both in canonical and pole gauge \cite{Yan93, Yan2011JGSP} and one can obtain $\tilde{\Lambda}$
for GMV from those as it was described in \cite{YanJMP2011}. The expressions for the recursion operators
for $\asl(n,\mathbb{C})$-CBC in pole gauge is also known \cite{YanVi2012SIGMA}. From it one also could easily
find $\tilde{\Lambda}_\pm$. Let us note that in \cite{GMVSIGMA} we already mentioned what is called
recursion operator is in fact  $\tilde{\Lambda}^2_{\pm}$ and it has been introduced as a product of
two operators which differ from $\tilde{\Lambda}_{\pm}$. This complicates the picture (we shall discuss
that a little further).

Let us continue with the calculation of the coefficients $\tilde{A}_{k}$. For the coefficient $\tilde{A}_{n-1}$
one has that $\rmi\partial_x\tilde{A}_n=[S,\tilde{A}_{n-1}]$ and since $\tilde{A}_n\in {\mathfrak{h}_S}$ there
are scalar functions $\alpha$, $\beta$ such that $\tilde{A}_n=\alpha S+\beta S_1$.
This gives
\[\rmi\alpha_x S+\rmi\beta_x S_1+\rm i\alpha S_x+\rm i\beta (S_1)_x\in \mathfrak{h}_S^{\perp}\]
and therefore $\alpha$ and $\beta$ are constants. Thus
\[\tilde{A}_{n-1}=\rm i\alpha \ad_S^{-1}S_x+\rmi\beta \ad_S^{-1}S_{1x}\]
and the hierarchy of NLEEs related to $\tilde{L}$ in general position
is
\begin{equation}\label{eq:EEZSSspole3a}
-\ad_{S}^{-1}\partial_t S+(\tilde{\Lambda}_{\pm})^n (\ad_S^{-1}(\alpha S_x+\beta (S_{1x})=0.
\end{equation}
In the case of ${\rm GMV_{\epsilon}}$ system $\tilde{A}_{n-1}\in \rmi\su(\epsilon)$ so $\alpha$ and $\beta$
must be real. Next, if $\tilde{A}_n\in {\mathfrak{h}}_1$ and $n=1 (\mod 2)$ because of the (\ref{eq:d1})
and (\ref{eq:d2}) we shall have automatically 
\begin{eqnarray}\label{eq:coher}
&&\tilde{A}_{2k}\in ({\mathfrak{h}}_1\oplus {\mathfrak{f}}_1)\cap \rm i\su(\epsilon),
\qquad \tilde{A}_{2k-1}\in ({\mathfrak{h}}_0\oplus {\mathfrak{f}}_0)\cap \rm i\su(\epsilon).
\end{eqnarray}
If $\tilde{A}_n\in {\mathfrak{h}}_0$ and $n=0 (\mod 2)$  we have again (\ref{eq:coher}). Thus the Mikhailov
type reductions are compatible with the action of recursion operator and the general form of the equations
related to the ${\rm GMV_{\epsilon}}$ system is:
\begin{eqnarray}\label{eq:NLEEs1}
&&\ad_{S}^{-1}\partial_t S=
\sum\limits_{k=0}^r a_{2k}(\tilde{\Lambda}_{\pm})^{2k} \ad_S^{-1}(S_x)+
\sum\limits_{k=1}^m a_{2k-1}(\tilde{\Lambda}_{\pm})^{2k-1} (\ad_S^{-1}(S_{1x})
\end{eqnarray}
where $a_i$ are some real constants. This is the hierarchy found in \cite{GMVSIGMA} although
it was presented there in a different form.

Let us denote the space of rapidly decreasing functions $\tilde{Z}(x)$ with values in $\mathfrak{f}_0$
(${\mathfrak{f}}_1$) by $\mathfrak{f}_0[x]$ (${\mathfrak{f}}_1[x]$). Since $\mathfrak{f}_0$ and $\mathfrak{f}_1$
are orthogonal with respect to the Killing form and taking into account Remark \ref{Rem:adS} we have
\begin{eqnarray}\nonumber
&&\tilde{Z}\in {\mathfrak{f}}_0[x],~\\ \label{eq:R1}
&& \tilde{\Lambda}_\pm(\tilde{Z})=\rmi {\ad}_S^{-1}
\displaystyle\left( \partial_x\tilde{Z} + \frac{S_{1x }}{4}\int\limits_{\pm\infty}^x
\langle \tilde{Z}, S_{1y} \rangle \rmd y \right)\in {\mathfrak{f}}_1[x],\\ \nonumber
&&\tilde{Z}\in {\mathfrak{f}}_1[x],\\  \label{eq:R2}
&&~ \tilde{\Lambda}_\pm(\tilde{Z})=\rmi {\ad}_S^{-1}
\displaystyle\left(\partial_x\tilde{Z} + \frac{S_x}{12}\int\limits_{\pm\infty}^x
\langle \tilde{Z}, S_y \rangle \rmd y \right)\in {\mathfrak{f}}_0[x] .
\end{eqnarray}
Now, let us do the following:
\begin{itemize}
\item Put in the above expressions $S=-L_1$ (in order to express everything through $L_1$).
\item Take into account  that what is called Killing form in \cite{GMVSIGMA} and denoted $\langle X,Y\rangle$
is not the canonical definition of the Killing form, it is in fact $\tr(XY)$ and as we know in our notation
$\langle X,Y\rangle=6\tr(XY)$.
\item Take into account that because $\tilde{Z}$ is orthogonal to ${\mathfrak{h}}_S$ we have $\langle \tilde{Z}(x),
S_x \rangle =-\langle \tilde{Z}_x, S(x) \rangle$ and $\langle \tilde{Z}(x), S_{1x} \rangle =-\langle \tilde{Z}_x, S_1(x) \rangle$. 
\end{itemize}
Then one could convince himself that  the operators $\Lambda^{\pm}_1,\Lambda^{\pm}_2$ introduced in \cite{GMVSIGMA}
are defined by the expressions standing in the right hand sides of (\ref{eq:R1}) and (\ref{eq:R2}), see \cite{YanJMP2011}.
In other words, $\Lambda^{\pm}_1,\Lambda^{\pm}_2$ are restrictions of the operators $\tilde{\Lambda}_{\pm}$ on
the spaces ${\mathfrak{f}}_0[x]$  and ${\mathfrak{f}}_1[x]$ respectively. If the hierarchy of the soliton
equations (\ref{eq:NLEEs1}) is written in terms of  $\Lambda^{\pm}_1,\Lambda^{\pm}_2$ it acquires the form
that was presented in \cite{GMVSIGMA}.

\section{Completeness Relations for the $\rm GMV_{\pm}$ System}

We have seen that the consideration of the NLEEs hierarchy (\ref{eq:NLEEs1}) shows that the operator `moving'
the equations along the hierarchy is $\tilde{\Lambda}_{\pm}^2$ and it does not depend on the real form. The
geometric interpretation of the hierarchies and their conservation laws which is developed for the case of
$\rm GMV_+$ \cite{YanVi2012JNMP} also suggests that the appropriate operator we must consider is
$\tilde{\Lambda}_{\pm}^2$. It remains to consider the third important aspect of the recursion operators
-- their relations to the so-called expansions over the adjoint solutions. In that approach one considers
these expansions and then finds operators, for which the adjoint solutions entering into them are eigenfunctions. 
The importance of the adjoint solutions is related to the fact that they form complete systems, so roughly
speaking, one can expand $S$ and its variation over them and then have the famous interpretation of the
inverse scattering method as a generalized Fourier transform, see \cite{IKhKi94,GerViYa2008}. Recently,
the theory of the expansions over adjoint solutions has been extended to the case when there are
reductions. In \cite{GYaSAM2015}, the theory for the CBC systems with $\mathbb{Z}_p$ reductions was
presented in full taking into account the discrete spectrum while in \cite{Yan2015SPT} is discussed
also the gauge-covariant formulation of the expansions with $\mathbb{Z}_p$ reductions. The geometric
aspects of the NLEEs associated with CBC systems in canonical and pole gauge with reductions were
presented recently in \cite{Yan2015GT}.

Our approach will be the following. We start from the spectral theory of the generating operators
for (\ref{eq:GZSours}) (a GZS system in canonical gauge) which is very well known and from it obtain
the theory of  the generating operators for $\rm GMV_{\pm}$ (a GZS system in pole gauge). So let us
sketch the theory for GZS in canonical gauge,  for all the details see  \cite{Yan93,GerViYa2008}.
The adjoint solutions for the system GZS system (\ref{eq:GZSours}) are defined as follows: 
\begin{equation}\label{eq:adjcan}
{\bf e}_\alpha^\pm := \pi_0\left( {\chi}^\pm E_\alpha
\hat{{\chi}}^\pm \right),\qquad \alpha\in \Delta
\end{equation}
where $\pi_0$ is the orthogonal projector on $\mathfrak{h}^{\perp}$ where $\mathfrak{h}$ is the
Cartan subalgebra of $\asl(3,\mathbb{C})$ and $\hat{\chi}$ stands for the matrix inverse to
$\chi\in SL(3,\mathbb{C})$.  The generating operators $\Lambda_{\pm}$ then have the form:
\begin{equation}\label{eq:}
\Lambda_\pm (Y(x)) = {\rm ad}_{J_0}^{-1}\, \left[\rmi \partial_x Y + \pi_0 {\rm ad}_q Y(x)
+ {\rm ad}_q \int\limits_{\pm\infty}^x ({\bf 1} -
\pi_0) {\rm ad}_q Y(y) dy \right] .
\end{equation}
As mentioned, the above formula together with its derivation could be found in many sources,
see for example \cite{G86,GYa94, GerViYa2008,GYaSAM2015}.  From the asymptotic behaviour of
the solutions $\chi^{\pm}$ follows that for $\alpha>0$ we have
\begin{equation}\label{eq:eigenadj}
\begin{array}{c}
{\Lambda}_{-}({\bf e}^{+}_{\alpha}(x,\lambda))=\lambda {\bf e}^{+}_{\alpha}(x,\lambda),
\qquad{\Lambda}_{-}({\bf e}^{-}_{-\alpha}(x,\lambda))=\lambda {\bf e}^{-}_{-\alpha}(x,\lambda)\\[4pt]
{\Lambda}_{+}({\bf e}^{+}_{-\alpha}(x,\lambda))=\lambda {\bf e}^{+}_{-\alpha}(x,\lambda),
\qquad {\Lambda}_{+}({\bf e}^{-}_{\alpha}(x,\lambda))=\lambda {\bf e}^{-}_{\alpha}(x,\lambda).
\end{array}
\end{equation}
(It is of no importance whether or not we have some reductions). Let us define now the adjoint solutions
for the $\rm GMV_{\pm}$ in the following way: 
\begin{equation}\label{eq:adjpole}
\tilde{\bf e}_\alpha^\pm := \pi_S\left( \tilde{\chi}^\pm E_\alpha
\hat{\tilde{\chi}}^\pm \right),\qquad \alpha\in \Delta.
\end{equation}
Let us remember now that we have a reduction defined by the automorphism $h$. Taking into account
that $h$ commutes with $\pi_S$ (see Proposition (\ref{prop:commpisigh})) and the properties of
$\psi_0$ (see (\ref{eq:sympsio}))  we see that $h(\tilde{\bf e}_\alpha^\pm)(x,\lambda)$ equals
\[\pi_S\left(H \tilde{\chi}^\pm E_\alpha
\hat{\tilde{\chi}}^\pm H\right)(x,\lambda)= \pi_S\left(\tilde{\chi}^\mp K E_\alpha K
\hat{\tilde{\chi}}^\mp\right)(x,-\lambda)= \pi_S\left(\tilde{\chi}^\mp E_{{\mathcal K}\alpha}
\hat{\tilde{\chi}}^\mp\right)(x,-\lambda)\]
so we have:
\begin{equation}\label{eq:Kadj}
h(\tilde{\bf e}_\alpha^\pm)(x,\lambda)=\tilde{\bf e}_{{\mathcal K}\alpha}^\mp(x,-\lambda).
\end{equation}
One proves easily that $\pi_S=\Ad(\psi^{-1}_0)\circ \pi_0\circ  \Ad(\psi_0)$, and then one sees that for
$\alpha\in \Delta$ we have that $\tilde{\bf e}^{\pm}_{\alpha}(x,\lambda)=\Ad(\psi^{-1}_0){\bf e}^{\pm}_{\alpha}(x,\lambda)$.
Here $\Ad(r)$ denotes the adjoint action of the group on its algebra, that is $\Ad(r)X=rXr^{-1}$. All
the theory then develops in a similar way as in the case when we make a gauge transformation with a Jost
solution which is very well known \cite{GYa94}. For example, one has that
\begin{equation}
\tilde{\Lambda}_{\pm}=\Ad(\psi^{-1}_0)\circ \Lambda_{\pm}\circ  \Ad(\psi_0).
\end{equation}
As a consequence, for $\alpha\in \Delta_+$ we have
\begin{equation}\label{eq:eigentilde}
\begin{array}{c}
\tilde{\Lambda}_{-}(\tilde{\bf e}^{+}_{\alpha}(x,\lambda))=\lambda \tilde{\bf e}^{+}_{\alpha}(x,\lambda),
\qquad \tilde{\Lambda}_{-}(\tilde{\bf e}^{-}_{-\alpha}(x,\lambda))=\lambda \tilde{\bf e}^{-}_{-\alpha}(x,\lambda),\\[4pt]
\tilde{\Lambda}_{+}(\tilde{\bf e}^{+}_{-\alpha}(x,\lambda))=\lambda \tilde{\bf e}^{+}_{-\alpha}(x,\lambda),
\qquad \tilde{\Lambda}_{+}(\tilde{\bf e}^{-}_{\alpha}(x,\lambda))=\lambda \tilde{\bf e}^{-}_{\alpha}(x,\lambda).
\end{array}
\end{equation}
We are now ready to discuss the completeness of the adjoint solutions for the GZS system in pole gauge
(which under reductions becomes our $\rm GMV_{\pm}$ problem). In order to simplify the formulas we shall
have further, let us adopt the following notation: for the functions $X(x), Y(x):\bbr \to\asl(3,\mathbb{C})$
we put
\begin{equation}
\langle\langle X,Y\rangle\rangle:=\int\limits_{-\infty}^{+\infty}\langle X(x), Y(x)\rangle \rmd x.
\end{equation}
First, we write the completeness relations for the 'adjoint` solutions of the canonical gauge GZS system
(details can be found in \cite{G86,GerViYa2008}). We also remind that we disregard the discrete spectrum.
Then for the GZS system in canonical gauge (\ref{eq:GZSours}) the following theorem holds:
\begin{theorem}\label{th:comprelcan}
Let ${\cal S}$ be the space of sufficiently smooth functions defined on the real axis with values in
$\mathfrak{h}^{\perp}$ tending fast enough to zero when $x\to\pm\infty$. Then for every ${Z}(x) \in {\cal S}$
the following expansion formulas hold ($\eta=\pm$):
\begin{eqnarray}\label{eq:expancan}
&&{Z}(x) = \displaystyle\frac{1}{2\pi} \displaystyle \int\limits_{-\infty}^\infty \left[
\sum\limits_{\alpha
\in \Delta_{+}} {\bf e}^\eta_\alpha(x,\lambda)
\langle\langle {\bf e}^\eta_{-\alpha},[J_0,{Z}]\rangle\rangle   -
{\bf e}^{-\eta}_{-\alpha}(x,\lambda)\langle\langle{\bf e}^{-\eta}_{\alpha},[J_0,{Z}]\rangle\rangle  \right] \rm d\lambda.
 \end{eqnarray}
\end{theorem}
Let us make now a gauge transformation in the above completeness relations for the GZS system in canonical
gauge, one will easily see that they transform to completeness relations for the GZS system in pole gauge
(which becomes $\rm GMV_{\pm}$ if we impose reductions): 
\begin{theorem}\label{th:comprel}
Let $\tilde{\cal S}$ be the space of sufficiently smooth functions defined on the real axis with values
in $\mathfrak{h}^{\perp}_S(x)$ tending fast enough to zero when $x\to\pm\infty$. Then for every
$\tilde{Z}(x) \in \tilde{\cal S}$ the following expansion formulas hold ($\eta=\pm$):
\begin{eqnarray}\label{eq:expan}
&&\tilde{Z}(x) = \displaystyle\frac{1}{2\pi} \displaystyle \int\limits_{-\infty}^\infty \left[
\sum\limits_{\alpha
\in \Delta_{+}} \tilde{\bf e}^\eta_\alpha(x,\lambda)
\langle\langle \tilde{\bf e}^\eta_{-\alpha},[S,\tilde{Z}]\rangle\rangle   -
\tilde{\bf e}^{-\eta}_{-\alpha}(x,\lambda)\langle\langle \tilde{\bf e}^{-\eta}_{\alpha},[S,\tilde{Z}]\rangle\rangle  \right] \rm d\lambda.
\end{eqnarray}
\end{theorem}
Note that we have two expansions here -- for $\eta=+$ and $\eta=-$. One is over the eigenfunctions of
$\tilde{\Lambda}_{-}$, the other is over the eigenfunctions of  $\tilde{\Lambda}_{+}$. In fact the above
is the spectral theorem for these operators. 

Now we assume that we have the reductions defined by $h$ and $\sigma_{\epsilon}$ and we are going to transform
these expansions so that it will be easier to see the action of $h$ and $\sigma_{\epsilon}$ on these expansions
when we expand functions that have values in the eigenspaces  $\mathfrak{f}_0$, $\mathfrak{f}_1$ respectively or
are `real' or `imaginary' with respect to the complex conjugation $\sigma_{\epsilon}$.

Let us start with the reduction defined by $h$. We have seen that $h$ acts on the adjoint solutions according
to (\ref{eq:Kadj}). Then using the invariance of the Killing form with respect to the action of $h$ we get
\begin{eqnarray*}
\int_{-\infty}^{+\infty}\langle \tilde{\bf e}_{-\alpha}^\eta(x, -\lambda), [S,\tilde{Z}]\rangle \rmd x=\int_{-\infty}^{+\infty}\langle h(\tilde{\bf e}_{-\alpha}^\eta(x, -\lambda)), h([S,\tilde{Z}])\rangle \rmd x\\
=\int_{-\infty}^{+\infty}\langle \tilde{\bf e}_{-{\cal K}\alpha}^{-\eta}(x, \lambda), H[S,\tilde{Z}]H\rangle \rmd x = -\int_{-\infty}^{+\infty}\langle \tilde{\bf e}_{-{\cal K}\alpha}^{-\eta}(x, \lambda), [S,h(\tilde{Z})]\rangle \rmd x .
\end{eqnarray*}
Therefore
\begin{eqnarray*}
\tilde{\bf e}^\eta_\alpha(x,-\lambda) \langle\langle \tilde{\bf e}^\eta_{-\alpha}(x,-\lambda),[S,\tilde{Z}]\rangle\rangle
= -h(\tilde{\bf e}^{-\eta}_{{\cal K}\alpha}(x,\lambda))\int_{-\infty}^{+\infty}\langle \tilde{\bf e}_{-{\cal K}\alpha}^{-\eta}(x, \lambda), [S,h(\tilde{Z})]\rangle \rmd x .
\end{eqnarray*}
Suppose for example $h(\tilde{Z})=\tilde{Z}$. Then the above is simply written as:
$$
\tilde{\bf e}^\eta_\alpha(x,-\lambda) \langle\langle \tilde{\bf e}_{-\alpha}^{\eta}(x, -\lambda), [S,\tilde{Z}]\rangle\rangle=
-h(\tilde{\bf e}^{-\eta}_{{\cal K}\alpha}(x,\lambda))\langle \langle \tilde{\bf e}_{-{\cal K}\alpha}^{-\eta}(x, \lambda), [S,\tilde{Z}]\rangle\rangle.
$$
Since $\mathcal{K}$ maps positive roots into negative and vice versa, we shall have:
$$
\begin{array}{l}
\sum_{\alpha\in \Delta_{+}}\tilde{\bf e}^\eta_\alpha(x,-\lambda) \langle\langle \tilde{\bf e}_{-\alpha}^{\eta}(x, -\lambda), [S,\tilde{Z}]\rangle\rangle=
-\sum_{\alpha\in \Delta_{+}} h(\tilde{\bf e}^{-\eta}_{-\alpha}(x,\lambda)) \langle\langle \tilde{\bf e}_{\alpha}^{-\eta}(x, \lambda), [S,\tilde{Z}]\rangle\rangle,\\[6pt]
\sum_{\alpha\in \Delta_{+}}\tilde{\bf e}^{-\eta}_{-\alpha}(x,-\lambda) \langle\langle \tilde{\bf e}_{\alpha}^{-\eta}(x, -\lambda), [S,\tilde{Z}]\rangle\rangle=
-\sum_{\alpha\in \Delta_{+}} h(\tilde{\bf e}^{\eta}_{\alpha}(x,\lambda)) \langle\langle \tilde{\bf e}_{-\alpha}^{\eta}(x, \lambda), [S,\tilde{Z}]\rangle\rangle.
\end{array}
$$
Now, let us make change of variables $\lambda\mapsto -\lambda$  in the integral standing in the right hand side of
(\ref{eq:expan}) and  use the above relations. We obtain:
\begin{eqnarray}
&&\quad\tilde{Z}(x) = \displaystyle\frac{1}{2\pi} \displaystyle \int\limits_{-\infty}^\infty \left[
\sum\limits_{\alpha
\in \Delta_{+}}h( \tilde{\bf e}^\eta_\alpha(x,\lambda))
\langle\langle \tilde{\bf e}^\eta_{-\alpha},[S,\tilde{Z}]\rangle\rangle   -
h(\tilde{\bf e}^{-\eta}_{-\alpha}(x,\lambda))\langle\langle \tilde{\bf e}^{-\eta}_{\alpha},[S,\tilde{Z}]\rangle\rangle  \right] \rmd\lambda.
 \end{eqnarray}
But then combining the original expansion with this one we immediately get:
\begin{eqnarray}\label{eq:prnewexp1}
&&\tilde{Z}(x) = \displaystyle\frac{1}{2\pi} \displaystyle \int\limits_{-\infty}^\infty \left[
\sum\limits_{\alpha
\in \Delta_{+}}\tilde{\bf s}^{\eta}_\alpha(x,\lambda)
\langle\langle \tilde{\bf e}^\eta_{-\alpha},[S,\tilde{Z}]\rangle\rangle   -
\tilde{\bf s}^{-\eta}_{-\alpha}(x,\lambda))\langle\langle \tilde{\bf e}^\eta_{\alpha},[S,\tilde{Z}]\rangle\rangle\right] \rmd\lambda
\end{eqnarray}
where
\begin{equation} \label{eq:sfun}
\tilde{\bf s}^{\eta}_{\pm \alpha}(x,\lambda)=\displaystyle \frac{1}{2}\left(\tilde{\bf e}^{\eta}_{\pm\alpha}(x,\lambda)+h(\tilde{\bf e}^{\eta}_{\pm\alpha}(x,\lambda))\right).
\end{equation}
In view of the formulas  we shall write a little further, we also note, that one has
$$
\langle\langle \tilde{\bf e}^\eta_{\mp \alpha},[S,\tilde{Z}]\rangle\rangle=\langle\langle h(\tilde{\bf e}^\eta_{\mp \alpha}),h([S,\tilde{Z}])\rangle\rangle=
$$
$$
-\langle\langle h (\tilde{\bf e}^\eta_{\mp \alpha}),[S,\tilde{Z}]\rangle\rangle
$$
so one can write
$$
\langle\langle \tilde{\bf e}^\eta_{\mp \alpha},[S,\tilde{Z}]\rangle\rangle=\langle\langle \tilde{\bf a}^\eta_{\mp \alpha},[S,\tilde{Z}]\rangle\rangle
$$
where
\begin{equation} \label{eq:afun}
\tilde{\bf a}^{\eta}_{\pm \alpha}(x,\lambda)=\displaystyle \frac{1}{2}\left(\tilde{\bf e}^{\eta}_{\pm\alpha}(x,\lambda)-h(\tilde{\bf e}^{\eta}_{\pm\alpha}(x,\lambda))\right).
\end{equation}
Thus finally we obtain:
\begin{eqnarray}\label{eq:newexpviaa}
&&\tilde{Z}(x) = \displaystyle\frac{1}{2\pi} \displaystyle \int\limits_{-\infty}^\infty \left[
\sum\limits_{\alpha
\in \Delta_{+}}\tilde{\bf s}^{\eta}_\alpha(x,\lambda)
\langle\langle \tilde{\bf a}^\eta_{-\alpha},[S,\tilde{Z}]\rangle\rangle   -
\tilde{\bf s}^{-\eta}_{-\alpha}(x,\lambda))\langle\langle \tilde{\bf a}^\eta_{\alpha},[S,\tilde{Z}]\rangle\rangle\right] \rmd\lambda.
\end{eqnarray}
If instead of $h(\tilde{Z})=\tilde{Z}$ we assume that $h(\tilde{Z})=-\tilde{Z}$ then in the same manner we shall obtain
\begin{eqnarray}\label{eq:newexpvias}
&&\tilde{Z}(x) = \displaystyle\frac{1}{2\pi} \displaystyle \int\limits_{-\infty}^\infty \left[
\sum\limits_{\alpha
\in \Delta_{+}}\tilde{\bf a}^{\eta}_\alpha(x,\lambda)
\langle\langle \tilde{\bf e}^\eta_{-\alpha},[S,\tilde{Z}]\rangle\rangle -
\tilde{\bf a}^{-\eta}_{-\alpha}(x,\lambda))\langle\langle \tilde{\bf e}^\eta_{\alpha},[S,\tilde{Z}]\rangle\rangle\right] \rmd\lambda \\ \nonumber
&& = \displaystyle\frac{1}{2\pi} \displaystyle \int\limits_{-\infty}^\infty \left[
\sum\limits_{\alpha
\in \Delta_{+}}\tilde{\bf a}^{\eta}_\alpha(x,\lambda)
\langle\langle \tilde{\bf s}^\eta_{-\alpha},[S,\tilde{Z}]\rangle\rangle -
\tilde{\bf a}^{-\eta}_{-\alpha}(x,\lambda))\langle\langle \tilde{\bf s}^\eta_{\alpha},[S,\tilde{Z}]\rangle\rangle\right] \rmd\lambda.
\end{eqnarray}
Naturally,
\begin{equation}
h(\tilde{\bf s}^{\eta}_{\pm \alpha}(x,\lambda))=\tilde{\bf s}^{\eta}_{\pm \alpha}(x,\lambda), \quad h(\tilde{\bf a}^{\eta}_{\pm \alpha}(x,\lambda))=-\tilde{\bf a}^{\eta}_{\pm \alpha}(x,\lambda).
\end{equation}  
Thus in case $h(\tilde{Z})=\tilde{Z}$ or $h(\tilde{Z})=-\tilde{Z}$ the expansions could be written in terms of
new sets of adjoint solutions, $\tilde{\bf s}^{\eta}_{\pm \alpha}(x,\lambda)$ in the first case and
$\tilde{\bf a}^{\eta}_{\pm \alpha}(x,\lambda)$ in the second, that reflect the symmetry of $\tilde{Z}$. 

Returning to the properties of the new families we introduced, we easily see that ${\bf s}^{\eta}_{\pm \alpha}(x,\lambda)$
and ${\bf a}^{\eta}_{\pm \alpha}(x,\lambda)$  are not eigenfunctions of $\tilde{\Lambda}_{-}$ or $\tilde{\Lambda}_{+}$.
In order to see what happens, we look how $\tilde{\Lambda}_{\pm}$ is related to $h$ and $\sigma_{\epsilon}$.
\begin{lemma}\label{lemma:ropvsred}
The following relations hold:  
\begin{equation}\label{eq:ropvsred}
h\circ \tilde{\Lambda}_{\pm} = - \tilde{\Lambda}_{\pm}\circ h, \quad \sigma_{\epsilon}\circ \tilde{\Lambda}_{\pm} = - \tilde{\Lambda}_{\pm}\circ \sigma_{\epsilon}.
\end{equation} 
\end{lemma}
\textit{Proof}. The proof of the first relation is straightforward, let us prove the second one. First, assume that $\sigma_{\epsilon}(\tilde{Z})=\tilde{Z}$ (The case when $\sigma_{\epsilon}(\tilde{Z})=-\tilde{Z}$ is treated in a similar way.)  First, we remark that $\sigma_{\epsilon}(S_x)=-S_x$ and that  $\sigma_{\epsilon}(S_1)=-S_1$, consequently  $\sigma_{\epsilon}(S_{1x})= - S_{1x}$. Next, we have
$$
\langle \tilde{Z}, S_x\rangle^*=\langle \sigma_{\epsilon}(\tilde{Z}), \sigma_{\epsilon}(S_x)\rangle=-\langle \tilde{Z}, S_x\rangle
$$
and similarly
$$
\langle \tilde{Z}, S_{1x}\rangle^*=\langle \sigma_{\epsilon}(\tilde{Z}), \sigma_{\epsilon}(S_{1x})\rangle=-\langle \tilde{Z}, S_{1x}\rangle.
$$
So in case $\sigma_{\epsilon}(\tilde{Z})=\tilde{Z}$ (we call $\tilde{Z}$ real in this case) the integrands in
the formula for recursion operator are purely imaginary while for $\sigma_{\epsilon}(\tilde{Z})=-\tilde{Z}$
(we then say that  $\tilde{Z}$ is purely imaginary) these expressions will be real. Then, taking into account
(\ref{eq:SymSs}) we get 
\begin{eqnarray}\nonumber
&&-\sigma_{\epsilon}\tilde{\Lambda}_\pm(\tilde{Z})=\\ \nonumber
&&\rmi {\ad}_S^{-1} \pi_S
\displaystyle\left(\partial_x(\sigma_{\epsilon}\tilde{Z}) + \frac{\sigma_{\epsilon}S_x}{12}\int\limits_{\pm\infty}^x \langle \tilde{Z}, S_y \rangle^* \rmd y +\frac{(\sigma_{\epsilon}S_{1x}) }{4}\int\limits_{\pm\infty}^x
\langle \tilde{Z}, S_{1y} \rangle^* \rmd y \right)=\\ \nonumber
&&\rmi {\ad}_S^{-1} \pi_S
\displaystyle\left(\partial_x(\tilde{Z}) + \frac{S_x}{12}\int\limits_{\pm\infty}^x \langle \tilde{Z}, S_y \rangle \rmd y +\frac{S_{1x }}{4}\int\limits_{\pm\infty}^x
\langle \tilde{Z}, S_{1y} \rangle\rmd y \right)=\\ \nonumber
&&\tilde{\Lambda}_\pm(\tilde{Z})=\tilde{\Lambda}_\pm(\sigma_{\epsilon}\tilde{Z}).
\end{eqnarray}
As mentioned, one can show in a similar way that when $\sigma_{\epsilon}(\tilde{Z})=-\tilde{Z}$  we
have the same relation. Finally, as each $\tilde{Z}$ could be written in a unique way as $\tilde{Z}=\tilde{Z}_1+\tilde{Z}_2$
where $\sigma_{\epsilon}(\tilde{Z}_1)=-\tilde{Z}_1$,   $\sigma_{\epsilon}(\tilde{Z}_2)=\tilde{Z}_2$ then
the second relation in (\ref{eq:ropvsred}) is proved for arbitrary  $\tilde{Z}$. 

Because of the relations (\ref{eq:eigentilde}) and the above lemma, for $\alpha\in \Delta_{+}$ one has
\begin{equation}\label{eq:asminus}
\begin{array}{c}
\tilde{\Lambda}_{-}(\tilde{\bf s}^{+}_{\alpha}(x,\lambda))=\lambda \tilde{\bf a}^{+}_{\alpha}(x,\lambda), \qquad \tilde{\Lambda}_{-}(\tilde{\bf s}^{-}_{-\alpha}(x,\lambda))=\lambda \tilde{\bf a}^{-}_{-\alpha}(x,\lambda),\\[4pt]
\tilde{\Lambda}_{-}(\tilde{\bf a}^{+}_{\alpha}(x,\lambda))=\lambda \tilde{\bf s}^{+}_{\alpha}(x,\lambda), \qquad \tilde{\Lambda}_{-}(\tilde{\bf a}^{-}_{-\alpha}(x,\lambda))=\lambda \tilde{\bf s}^{-}_{- \alpha}(x,\lambda),
\end{array}
\end{equation}
\begin{equation}\label{eq:asplus}
\begin{array}{c}
\tilde{\Lambda}_{+}(\tilde{\bf s}^{+}_{-\alpha}(x,\lambda))=\lambda \tilde{\bf a}^{+}_{-\alpha}(x,\lambda) , \qquad \tilde{\Lambda}_{+}(\tilde{\bf s}^{-}_{\alpha}(x,\lambda))=\lambda \tilde{\bf a}^{-}_{\alpha}(x,\lambda), \\[4pt]
\tilde{\Lambda}_{+}(\tilde{\bf a}^{+}_{-\alpha}(x,\lambda))=\lambda \tilde{\bf s}^{+}_{-\alpha}(x,\lambda) , \qquad \tilde{\Lambda}_{+}(\tilde{\bf a}^{-}_{\alpha}(x,\lambda))=\lambda \tilde{\bf s}^{-}_{\alpha}(x,\lambda).
\end{array}
\end{equation}
and one sees that the functions in the expansions when we have some symmetry with respect to $h$ are eigenfunctions for  $\tilde{\Lambda}^2_{-}$ ($\tilde{\Lambda}^2_{+}$) with eigenvalue $\lambda^2$. This together with the fact that when recursively finding the coefficients for the pencil (\ref{eq:pLax}) one effectively uses $\tilde{\Lambda}^2_{+}$ has led to the interpretation that in case we have $\mathbb{Z}_2$ reduction defined by $h$ the role of the Generating Operator is played by $\tilde{\Lambda}_{\pm}^2$. 

Let us consider now the other reduction that we have in the ${\rm GMV_{\pm}}$ system, the one defined by the complex conjugation $\sigma_{\epsilon}$. In accordance with Theorem \ref{fas_sym} the FAS fulfill $Q_{\epsilon}(\tilde{\chi}^{\mp}(x,\lambda))^{\dag}Q_{\epsilon}=(\tilde{\chi}^{\pm}(x,\lambda^*))^{-1}$ so on the real axis $Q_{\epsilon}(\tilde{\chi}^{\mp}(x,\lambda))^{\dag}Q_{\epsilon}=(\tilde{\chi}^{\pm}(x,\lambda^*))^{-1}$. Then taking into account Proposition (\ref{prop:commpisigh}) for $\alpha\in \Delta$
$$
Q_{\epsilon}\tilde{\bf e}_{\alpha}^{\eta}(x,\lambda)^{\dag}Q_{\epsilon}=\pi_S\tilde{\chi}^{-\eta}(x,\lambda) Q_{\epsilon}E^{\dag}_{\alpha}Q_{\epsilon}\hat{\tilde{\chi}}^{-\eta}(x,\lambda^*)
$$
and we need to take into account how $X\mapsto \sigma_{\epsilon}(X) = -Q_{\epsilon}X^{\dag}Q_{\epsilon}$ acts on the Cartan-Weil basis. One easily sees that
\begin{equation}
\begin{array}{l}
\sigma_{+}E_{\pm\alpha_1}=-E_{\mp\alpha_1},\qquad  \sigma_{+}E_{\pm\alpha_2}=-E_{\mp\alpha_2}, \qquad  \sigma_{+}E_{\pm\alpha_3}=-E_{\mp\alpha_3}\\[4pt]
\sigma_{-}E_{\pm\alpha_1}= E_{\mp\alpha_1},\qquad  \sigma_{-}E_{\pm\alpha_2}=E_{\mp\alpha_2}, \qquad  \sigma_{-}E_{\pm\alpha_3}=-E_{\mp\alpha_3}\\[4pt]
\sigma_{\epsilon} H_{\alpha_j} =  H_{\alpha_j}, \qquad j=1,2.
\end{array}
\end{equation}
In order to write down these relations in a concise way, we introduce the symbol $q_{\epsilon}(\alpha)$ which is equal to $-1$ if $\epsilon=1$ for arbitrary $\alpha\in \Delta$ but if $\epsilon=-1$ then $q_{\epsilon}(\alpha)=1$ for $\alpha=\pm\alpha_1,\pm\alpha_2$ and $q_{\epsilon}(\pm \alpha_3)= - 1$. Then $q_{\epsilon}(\alpha)q_{\epsilon}(-\alpha)=1$ and we have
\begin{equation}
\sigma_{\epsilon}E_{\alpha}=q_{\epsilon}(\alpha) E_{-\alpha}, \qquad \sigma_{\epsilon} H_{\alpha_i} = - H_{\alpha_i}, \qquad i=1,2.
\end{equation}
With the above notation $Q_{\epsilon}\tilde{\bf e}_{\alpha}^{\eta}(x,\lambda)^{\dag}Q_{\epsilon}=-q_{\epsilon}(\alpha)\tilde{\bf e}_{-\alpha}^{-\eta}(x,\lambda)$ or if one prefers expression through the complex conjugation
\begin{equation}
\sigma_{\epsilon}(\tilde{\bf e}_{\alpha}^{\eta}(x,\lambda))=q_{\epsilon}(\alpha)\tilde{\bf e}_{-\alpha}^{-\eta}(x,\lambda^*).
\end{equation}
Naturally, we have
\begin{equation}
\sigma_{\epsilon}(\tilde{\bf e}_{\alpha}^{\eta}(x,\lambda))=q_{\epsilon}(\alpha)\tilde{\bf e}_{-\alpha}^{-\eta}(x,\lambda),
\qquad \lambda\in \bbr.
\end{equation}
Consider now the coefficients $\langle\langle  \tilde{\bf e}^\eta_{-\alpha}, [S,\tilde{Z}]\rangle\rangle$ introduced in
the above and  and let us assume that $\sigma_{\epsilon}\tilde{Z}=\tilde{Z}$. We have
$$
\langle\langle  \tilde{\bf e}^\eta_{-\alpha}, [S,\tilde{Z}]\rangle\rangle=\langle\langle  \sigma_{\epsilon}\tilde{\bf e}^\eta_{-\alpha},  \sigma_{\epsilon}[S,\tilde{Z}]\rangle\rangle^*=-q_{\epsilon}(\alpha)\langle\langle  \tilde{\bf e}^{-\eta}_{\alpha}, [S,\tilde{Z}]\rangle\rangle^*.
$$
Consequently:
\begin{equation}
\sigma_{\epsilon}(\langle\langle  \tilde{\bf e}^\eta_{-\alpha}(\lambda), [S,\tilde{Z}]\rangle\rangle\tilde{\bf e}_{\alpha}^{\eta}(x,\lambda))=-\langle\langle  \tilde{\bf e}^{-\eta}_{\alpha}(\lambda), [S,\tilde{Z}]\rangle\rangle\tilde{\bf e}_{-\alpha}^{-\eta}(x,\lambda).
\end{equation}
In case $\sigma_{\epsilon}\tilde{Z}=-\tilde{Z}$ we obtain in a similar way 
\begin{equation}
\sigma_{\epsilon}(\langle\langle  \tilde{\bf e}^\eta_{-\alpha}(\lambda), [S,\tilde{Z}]\rangle\rangle\tilde{\bf e}_{\alpha}^{\eta}(x,\lambda))
= \langle\langle  \tilde{\bf e}^{-\eta}_{\alpha}(\lambda), [S,\tilde{Z}]\rangle\rangle\tilde{\bf e}_{-\alpha}^{-\eta}(x,\lambda).
\end{equation}
One could also see that in case $\sigma_{\epsilon}\tilde{Z}=\tilde{Z}$ we have
$$
\langle\langle  \tilde{\bf e}^\eta_{-\alpha}(\lambda), [S,\tilde{Z}]\rangle\rangle^*=\langle\langle  \sigma_{\epsilon}\tilde{\bf e}^\eta_{-\alpha}(\lambda), \sigma_{\epsilon}[S,\tilde{Z}]\rangle\rangle=-\langle\langle  \sigma_{\epsilon}\tilde{\bf e}^{\eta}_{-\alpha}(\lambda), [S,\tilde{Z}]\rangle\rangle.
$$
Thus when $\sigma_{\epsilon}\tilde{Z}=\tilde{Z}$ (respectively $\sigma_{\epsilon}\tilde{Z}=-\tilde{Z}$) the expansions
(\ref{eq:expan}) acquire the form:
\begin{equation}
\begin{array}{c}
\tilde{Z}(x) = \displaystyle\frac{1}{2\pi} \displaystyle \int\limits_{-\infty}^\infty \left[
\sum\limits_{\alpha
\in \Delta_{+}} (\mathbf{1}+\sigma_{\epsilon})\left( \tilde{\bf e}^\eta_\alpha(x,\lambda)
\langle\langle  \tilde{\bf e}^\eta_{-\alpha}(\lambda), [S,\tilde{Z}]\rangle\rangle \right) \right] \rmd\lambda
\end{array}
\end{equation}
and 
\begin{equation}
\begin{array}{c}
\tilde{Z}(x) = \displaystyle\frac{1}{2\pi} \displaystyle \int\limits_{-\infty}^\infty \left[
\sum\limits_{\alpha
\in \Delta_{+}} (\mathbf{1}-\sigma_{\epsilon})\left( \tilde{\bf e}^\eta_\alpha(x,\lambda)\langle\langle  \tilde{\bf e}^\eta_{-\alpha}(\lambda), [S,\tilde{Z}]\rangle\rangle\right) \right] \rmd\lambda
\end{array}
\end{equation}
respectively.

In order to obtain expansions when both reductions are present, we notice that since $\sigma_{\epsilon}$ and $h$
commute, in case  $\sigma_{\epsilon}\tilde{Z}=\tilde{Z}$ one has:
\begin{equation}
\begin{array}{l}
\sigma_{\epsilon}(\langle\langle  \tilde{\bf e}^\eta_{-\alpha}, [S,\tilde{Z}]\rangle\rangle\tilde{\bf s}_{\alpha}^{\eta}(x,\lambda))=-\langle\langle  \tilde{\bf e}^{-\eta}_{\alpha}, [S,\tilde{Z}]\rangle\rangle\tilde{\bf s}_{-\alpha}^{-\eta}(x,\lambda), \\[6pt]
\sigma_{\epsilon}(\langle\langle  \tilde{\bf e}^\eta_{-\alpha}, [S,\tilde{Z}]\rangle\rangle\tilde{\bf a}_{\alpha}^{\eta}(x,\lambda))=-\langle\langle  \tilde{\bf e}^{-\eta}_{\alpha}, [S,\tilde{Z}]\rangle\rangle\tilde{\bf a}_{-\alpha}^{-\eta}(x,\lambda)
\end{array}
\end{equation}
and in case $\sigma_{\epsilon}\tilde{Z}=-\tilde{Z}$ we have
\begin{equation}
\begin{array}{l}
\sigma_{\epsilon}(\langle\langle  \tilde{\bf e}^\eta_{-\alpha}, [S,\tilde{Z}]\rangle\rangle\tilde{\bf s}_{\alpha}^{\eta}(x,\lambda))=\langle\langle  \tilde{\bf e}^{-\eta}_{\alpha}, [S,\tilde{Z}]\rangle\rangle\tilde{\bf s}_{-\alpha}^{-\eta}(x,\lambda), \\[6pt]
\sigma_{\epsilon}(\langle\langle  \tilde{\bf e}^\eta_{-\alpha}, [S,\tilde{Z}]\rangle\rangle\tilde{\bf a}_{\alpha}^{\eta}(x,\lambda))=\langle\langle  \tilde{\bf e}^{-\eta}_{\alpha}, [S,\tilde{Z}]\rangle\rangle\tilde{\bf a}_{-\alpha}^{-\eta}(x,\lambda).
\end{array}
\end{equation}
Then we obtain:
\begin{theorem}\label{Th:exphsig}
The following expansions hold:
\begin{itemize}
\item In case  $\sigma_{\epsilon}\tilde{Z}=\tilde{Z}$,  $h\tilde{Z}=\tilde{Z}$
\begin{eqnarray}
&&\tilde{Z}(x) = \displaystyle\frac{1}{4\pi} \displaystyle \int\limits_{-\infty}^\infty \left[
\sum\limits_{\alpha
\in \Delta_{+}} (\mathbf{1}+\sigma_{\epsilon})\left( \tilde{\bf s}^\eta_\alpha(x,\lambda)
\langle\langle  \tilde{\bf e}^\eta_{-\alpha}, [S,\tilde{Z}]\rangle\rangle(\lambda) \right) \right] \rmd\lambda.
\end{eqnarray}
\item In case  $\sigma_{\epsilon}\tilde{Z}=\tilde{Z}$,  $h\tilde{Z}=-\tilde{Z}$
\begin{eqnarray}
&&\tilde{Z}(x) =\displaystyle\frac{1}{4\pi} \displaystyle \int\limits_{-\infty}^\infty \left[
\sum\limits_{\alpha
\in \Delta_{+}} (\mathbf{1}+\sigma_{\epsilon})\left( \tilde{\bf a}^\eta_\alpha(x,\lambda)
\langle\langle  \tilde{\bf e}^\eta_{-\alpha}, [S,\tilde{Z}]\rangle\rangle(\lambda) \right) \right] \rmd\lambda.
\end{eqnarray}

\item In case  $\sigma_{\epsilon}\tilde{Z}=-\tilde{Z}$,  $h\tilde{Z}=\tilde{Z}$
\begin{eqnarray}
&&\tilde{Z}(x) = \displaystyle\frac{1}{4\pi} \displaystyle \int\limits_{-\infty}^\infty \left[
\sum\limits_{\alpha
\in \Delta_{+}} (\mathbf{1}-\sigma_{\epsilon})\left( \tilde{\bf s}^\eta_\alpha(x,\lambda)
\langle\langle  \tilde{\bf e}^\eta_{-\alpha}, [S,\tilde{Z}]\rangle\rangle(\lambda) \right) \right] \rmd\lambda.
\end{eqnarray}
\item In case  $\sigma_{\epsilon}\tilde{Z}=-\tilde{Z}$,  $h\tilde{Z}=-\tilde{Z}$
\begin{eqnarray}
&&\tilde{Z}(x) = \displaystyle\frac{1}{4\pi} \displaystyle \int\limits_{-\infty}^\infty \left[
\sum\limits_{\alpha
\in \Delta_{+}} (\mathbf{1}-\sigma_{\epsilon})\left( \tilde{\bf a}^\eta_\alpha(x,\lambda)
\langle\langle  \tilde{\bf e}^\eta_{-\alpha}, [S,\tilde{Z}]\rangle\rangle(\lambda) \right) \right] \rmd\lambda.
\end{eqnarray}
\end{itemize}
\end{theorem}
\begin{corollary}
In view of Lemma \ref{lemma:ropvsred}, we see that the functions that stay in the integrands of the above expressions, that is  $(\mathbf{1}\pm \sigma_{\epsilon})\tilde{\bf s}^\eta_\alpha(x,\lambda)$ and $(\mathbf{1}\pm \sigma_{\epsilon})\tilde{\bf a}^\eta_\alpha(x,\lambda)$ remain eigenfunctions of $\tilde{\Lambda}_{\pm}^2$ with eigenvalues $\lambda^2$, that is, the reduction defined by a real form does not change the recursion operators.
\end{corollary}
Let us make some final comments about the expansions we obtained.  In the case $\rm GMV_{+}$ the families $\tilde{\bf s}^{\eta}_{\pm \alpha}(x,\lambda)$ and $\tilde{\bf a}^{\eta}_{\pm \alpha}(x,\lambda)$ as well as the relations (\ref{eq:asminus}), (\ref{eq:asplus}) were introduced in \cite{GMVSIGMA}. However, they were written in terms of the restrictions of $\tilde{\Lambda}_{\pm}$ on the spaces ${\mathfrak{f}}_0[x]$ (${\mathfrak{f}}_1[x]$), namely through the operators $\Lambda_{1}^{\pm}$, ${\Lambda}_{2}^{\pm}$ we mentioned already. This complicates their form and obscures their meaning. We believe that the form in which we cast them now and in relation to Lemma \ref{lemma:ropvsred} they are much easier to understand. As about the expansions (\ref{eq:newexpviaa}),  (\ref{eq:newexpvias}) and about expansions in Theorem \ref{Th:exphsig} they were not presented until now in their general form but only for the specific cases of expansions of the functions $S_x$ and $(S_1)_x$. Moreover, since in  \cite{GMVSIGMA} the theory of the $\rm GMV_{+}$ have been developed not in parallel (though by analogy) to the corresponding system in canonical gauge, the knowledge about the GZS system in canonical gauge could not be exploited in full and everything ought to be developed from the beginning. In particular, there have been considered more restricted boundary conditions than necessary, namely the case $u_{\pm}=0$, $v_{\pm}=\exp{\rmi\Phi_{\pm}}$. Consequently, all the results were obtained in less generality and with more effort.

The above expansions have big theoretical  interest. They show that similar to the general position case (without reductions) the 'potential' and its variation could be expanded over a family of functions that are eigenfunctions of the operator that appears when one recursively solves the equations for the coefficients $\tilde{A}_k$.  Of course, it is important also to know that when we have reductions the expansions could be written in a different form. Practically however, it is easier to adopt the following attitude. We could simply 'forget' that we have reductions. In other words, the whole Gauge Covariant Theory of the Generating Operators will proceed in the usual way, with or without reductions. Basically, all the formulas will remain true, one must simply take into account the following:
\begin{itemize}
\item Some additional symmetries of the scattering data (see the next Section) which lead to symmetries in the corresponding Riemann-Hilbert problem when one uses it for finding exact solutions.  
\item The operators that permit to `move' along the hierarchies are $\tilde{\Lambda}_{\pm}^2$.
\item Some of the conservation laws and Poisson structures in the corresponding hierarchies trivialize.
\end{itemize}
For the first of these items one can see \cite{GMVSIGMA}, as for the second and the third, according to the general theory (see  \cite{GerViYa2008} or \cite{Yan93} specifically for the case of $\asl(3,\mathbb{C})$) we have:
\begin{itemize}
\item The NLEEs (\ref{eq:NLEEs1}) have the following series of conservation laws:
\begin{equation}\label{eq:conslaws}
D_B^{(s)} = \displaystyle \int \limits_{-\infty}^\infty \left[ \int\limits_{-\infty}^y \langle S_y, \tilde{\Lambda}_{\pm}^{s}(\ad_S^{-1}\tilde{B}_y) \rangle  \rmd y \right] \rmd x, \qquad B\in \mathfrak{h}, \quad B=\const,\qquad \tilde{B}=gBg^{-1}
\end{equation}
where $s=1,2,\ldots$ Now, if $k(B)=-B$, then $\ad_S^{-1}\tilde{B}_x\in \mathfrak{h}_0$ and if $s$ is even
$\tilde{\Lambda}_{\pm}^{s}(\tilde{B}_x )\in \mathfrak{h}_0[x]$. Since $S_y\in \mathfrak{h}_1[x]$ we have
$D_B^{(s)}=0$. This of course does not happen if $s$ is odd. Analogously, $D_B^{(s)}=0$ if  $s$ is odd and
$k(B)=B$. Thus indeed some of the conservation laws trivialize.

\item The NLEEs  (\ref{eq:NLEEs1}) are Hamiltonian with respect to the hierarchy of symplectic forms:
\begin{equation}\label{eq:symplstrs}
\tilde{\Omega}^{(p)}(\delta S_1, \delta S_2)  = \int\limits_{-\infty}^\infty
\langle  \delta S_1, \tilde{\Lambda}^p\, {\rm ad}_S^{-1} \delta S_2
\rangle \rm d x,~~ p=0,1,2,\ldots
\end{equation}
where $\delta_1S, \delta_2S$ are some variations of $S$ and
\begin{equation}\label{3.33}
\tilde{\Lambda} =\frac{1}{2} (\tilde{\Lambda}_{+} +
\tilde{\Lambda}_{-}) .
\end{equation}
Now, since $\langle S,S\rangle=12$, $\delta_1S, \delta_2S$ are orthogonal to $S$ and since $h(S)=-S$ we have $h(\delta_1S)=-\delta_1S$,  $h(\delta_1S)=-\delta_1S$ so finally  $\delta_1S, \delta_2S\in \mathfrak{f}_1[x]$. Further,
$\ad_S^{-1}\delta_2S\in \mathfrak{f}_0[x]$ and therefore, if $p$ is even $\tilde{\Omega}^{(p)}$ is identically zero. 
The simplest case is of course the case $p=0$, that corresponds to the Kirillov Poisson structure $\ad_S$ (restricted to the manifold of the potentials) which obviously trivializes. On the contrary, for $p$ odd, these structures are not trivial. The symplectic structure for $p=1$ corresponds to the Poisson tensor $\rm i\partial_x$  (restricted to the manifold of the potentials) and does not trivialize, see \cite{YanVi2012JNMP} for the details.
\end{itemize}

\section{Application. Scattering Data}

As an application we are going to discuss the implications of the reduction over the scattering data related
to the $\asl(3,\mathbb{C})$-GZS system in canonical and in pole gauge. First of all, as well-known from the
theory of the CBC (GZS) systems, there are essentially two types of scattering data: a) associated with the
asymptotic of FAS b) associated with the coefficients in the expansions over the adjoint solutions. All
these sets of scattering data are equivalent, but we are not going to enter into this issue here.

\subsection{Scattering Data Related to the Asymptotics}

As it follows from the general theory of CBC (GZS) systems, the functions $m^{\pm}(x,\lambda)$ through
which we introduced the FAS for (\ref{eq:GZSours}) in case of real $\lambda$ have asymptotics, see for
example \cite{G86} or \cite{Yan93}:
\begin{equation}\label{eq:asipm}
\begin{array}{lr}
m^+(x,\lambda)=\rme^{\rmi\lambda J_0x} \chi^+(x,\lambda)\to S^{+} (\lambda),\qquad &x\to-\infty,\\[4pt]
m^+(x,\lambda)=\rme^{\rmi\lambda J_0x}\chi^+(x,\lambda) \to T^{-}(\lambda) D^{+}(\lambda),\qquad &x\to+\infty,\\[4pt]
m^-(x,\lambda)=\rme^{\rmi\lambda J_0x}\chi^-(x,\lambda) \to S^{-} (\lambda),\qquad &x\to-\infty,\\[4pt]
m^-(x,\lambda)=\rme^{\rmi\lambda J_0x}\chi^-(x,\lambda) \to T^{+} (\lambda)D^{-}(\lambda),\qquad &x\to+\infty.
\end{array}
\end{equation}
The matrices $S^+, T^+$ are upper triangular with with diagonal elements equal to $1$,  the matrices
$S^-,T^-$ are lower triangular with with diagonal elements equal to $1$ and the matrices $D^{\pm}$ are
diagonal with determinant $1$. Besides, $D^+(\lambda)$ could be defined in the upper half-plane $\mathbb{C}_+$
while $D^-(\lambda)$ could be defined  in the lower half-plane $\mathbb{C}_-$. They are meromorphic and the
poles correspond to the discrete spectrum. As agreed, we shall assume that there is no discrete spectrum,
so $D^+(\lambda)$ ($D^-(\lambda)$) are analytic in $\mathbb{C}_+$ ($\mathbb{C}_-$) respectively.

Since on the real line both $\chi^{\pm}(x,\lambda)$ exist, there is a non-degenerate matrix $R(\lambda)$ such that $\chi^{+}(x,\lambda)=\chi^{-}(x,\lambda)R(\lambda)$ and we get that
$$
R(\lambda)=\widehat{S}^{-}(\lambda){S}^{+}(\lambda)=
\widehat{D}^{-}(\lambda)\widehat{T}^{+}(\lambda){T}^{-}(\lambda){D}^{+}(\lambda),\qquad \lambda \in \bbr .
$$
One could cast this in an equivalent form, introducing a matrix $T(\lambda)$
\begin{equation}\label{eq:Gauss-tr}
T(\lambda) = T^{-}(\lambda) D^{+}(\lambda)
\widehat{S}^{+}(\lambda) = T^{+}(\lambda) D^{-}(\lambda)
\widehat{S}^{-}(\lambda), \qquad \lambda \in \bbr .
\end{equation}
The matrices $S^+, T^+$ are upper-triangular with diagonal elements equal to $1$, the matrices $S^-, T^-$
are lower triangular with diagonal elements equal to $1$ and the matrices $D^{\pm}$ are diagonal. The above
shows that the matrices we just introduced are factors of two Gauss decompositions of  the matrix $T(\lambda)$,
which is called the transition matrix. We have the following representations of the Gauss factors:
\begin{equation} \label{eq:stdGaussfact}
\begin{array}{l}
S^{+}(\lambda)=\displaystyle \exp \sum_{\alpha \in \Delta_{+}}
s^{+}_{\alpha}(\lambda)E_\alpha,   \qquad S^{-}(\lambda) = \exp
\sum_{\alpha \in \Delta_{+}}
s^{-}_{-\alpha}(\lambda)E_{-\alpha} , \\
D^{+}(\lambda) = \exp  \sum\limits_{i=1}^2{d}^{+}_i(\lambda)H_i , \qquad 
D^{-}(\lambda) = \exp  \sum\limits_{i=1}^2{d}^{-}_i(\lambda)H_i , \\[12pt]
T^{+}(\lambda) =  \exp \sum_{\alpha \in \Delta_{+}}
t^{+}_{\alpha}(\lambda)E_\alpha ,  \qquad T^{-}(\lambda) =  \exp
\sum_{\alpha \in \Delta_{+}} t^{-}_{-\alpha}(\lambda)E_{-\alpha}  .
\end{array}
\end{equation}
As it is known, see \cite{G86, GYa94, GerViYa2008} each one of the families of functions
\begin{equation}
\begin{array}{l}
{\cal F}_s=\{s_{\alpha}^+(\lambda), s_{-\alpha}^{-}(\lambda),\quad  \lambda\in \bbr,\quad \alpha\in\Delta_{+}\},\\[6pt]
{\cal F}_t=\{t_{\alpha}^+(\lambda), t_{-\alpha}^{-}(\lambda),\quad  \lambda\in \mathbb{R}, \quad\alpha\in\Delta_{+}\}
\end{array}
\end{equation}
could be taken as a scattering data, from which one could reconstruct the potential $q$ (and hence the potential function $S$). We have reductions and they impose some restrictions on the scattering data.  To start with, since $K\chi^{\pm}(x,\lambda)K=\chi^{\mp}(x,-\lambda)$, we get that
\begin{equation}\label{eq:STDsymm}
\begin{array}{l}
KS^+(-\lambda)K=S^-(\lambda), \qquad  KT^{-}(-\lambda)K=T^{+}(\lambda),\\[4pt] KD^+(-\lambda)K=D^{-}(\lambda).
\end{array}
\end{equation}
Using the quantities introduced in (\ref{eq:stdGaussfact}), the above relations could be written in an equivalent
form: for $\alpha\in \Delta_{+}$
\begin{equation}\label{eq:stdsymme}
\begin{array}{l}
s^{-}_{-\alpha}(-\lambda)=s^{+}_{-{\cal K}\alpha}(\lambda),\qquad  t^-_{{-\alpha}}(-\lambda)=t^{+}_{-{\cal K}\alpha}(\lambda),\\[4pt]
s^+_{{\alpha}}(-\lambda)=s^{-}_{{\cal K}\alpha}(\lambda),\qquad  t^+_{{\alpha}}(-\lambda)=t^{-}_{{\cal K}\alpha}(\lambda),\\[4pt]
d_{1}^+(-\lambda)=-d_2^{-}(\lambda),\qquad d_{2}^+(-\lambda)=-d_1^{-}(\lambda)
\end{array}
\end{equation}
where ${\cal K}$ is the action of the automorphism $K$ on the roots. The second involution leads to the relations:
\begin{eqnarray}
&&Q_{\epsilon}(S^{+}(\lambda))^{\dag}Q_{\epsilon}=(S^{-}(\lambda))^{-1},
\qquad Q_{\epsilon}(T^{+}(\lambda))^{\dag}Q_{\epsilon}=(T^{-}(\lambda))^{-1},\\
\nonumber 
&&(D^{+}(\lambda))^{\dag}=(D^{-}(\lambda))^{-1},\\
&&(s^+_{\alpha}(\lambda))^*=q_{\epsilon}(\alpha)s^{-}_{-\alpha}(\lambda),
\qquad  (t^+_{\alpha}(\lambda))^*=q_{\epsilon}(\alpha)t^{-}_{-\alpha}(\lambda),
\qquad \alpha\in \Delta_{+}\\ \nonumber
&&(d_{i}^+(\lambda))^*= -d_i^{-}(\lambda), \qquad i=1,2
\end{eqnarray}
where in all the above relations $\lambda$ is real. 

Now it is not hard to find the corresponding relations for the $\rm GMV_{\pm}$ system. Indeed, because
of (\ref{eq:FAStilde}) we get
\begin{equation}\label{eq:asipmtilde}
\begin{array}{lr}
\rme^{\rmi\lambda J_0x}g_{-}^{-1}\tilde{\chi}^+(x,\lambda) \to S^{+} (\lambda),\qquad &x\to-\infty,\\[4pt]
\rme^{\rmi\lambda J_0x}g_+^{-1}\tilde{\chi}^+(x,\lambda) \to \rme^{\rmi\Phi J'}T^{-}(\lambda) D^{+}(\lambda),
\qquad &x\to+\infty,\\[4pt]
\rme^{\rmi\lambda J_0x}g_-^{-1}\tilde{\chi}^-(x,\lambda) \to S^{-} (\lambda),\qquad &x\to-\infty,\\[4pt]
\rme^{\rmi\lambda J_0x}g_+^{-1}{\tilde\chi}^-(x,\lambda) \to \rme^{\rmi\Phi J'} T^{+} (\lambda)D^{-}(\lambda),
\qquad &x\to +\infty.
\end{array}
\end{equation}
Since $J'=H_1-H_2$ we easily get that
\begin{equation}
\begin{array}{l}
\rme^{\rmi\Phi J'}T^{-}(\lambda) D^{+}(\lambda)=\tilde{T}^{-}(\lambda) \tilde{D}^{+}(\lambda),\\[4pt]
\rme^{\rmi\Phi J'} T^{+} (\lambda)D^{-}(\lambda)= \tilde{T}^{+} (\lambda)\tilde{D}^{-}(\lambda)
\end{array}
\end{equation}
where 
\begin{equation}
\begin{array}{l}
\tilde{D}^{+}(\lambda)=\rme^{\rmi\Phi J'}{D}^{+}(\lambda), \qquad \tilde{D}^{-}(\lambda)=\rme^{\rmi\Phi J'}{D}^{-}(\lambda),\\[4pt]
\tilde{T}^{\pm}(\lambda)=\rme^{\rmi\Phi J'}{T}^{\pm}(\lambda)\rme^{-\rmi\Phi J'}.
\end{array}
\end{equation}
We also put $\tilde{S}^{\pm}(\lambda)=S^{\pm}(\lambda)$.  Since $KJ'K=J$ the factors $\tilde{D}^{\pm}(\lambda), \tilde{S}^{\pm}(\lambda),\tilde{T}^{+}(\lambda)$ obey the same relations (\ref{eq:STDsymm}) as the factors ${D}^{\pm}(\lambda), {S}^{\pm}(\lambda),{T}^{+}(\lambda)$. Also, we can write
\begin{equation} \label{eq:stdGaussfactpole}
\begin{array}{l}
\tilde{D}^{+}(\lambda) = \exp  \sum\limits_{i=1}^2\tilde{d}^{+}_i(\lambda)H_i , \qquad 
\tilde{D}^{-}(\lambda) = \exp  \sum\limits_{i=1}^2\tilde{d}^{-}_i(\lambda)H_i  \\[8pt]
\tilde{T}^{\pm}(\lambda) =  \exp \sum_{\alpha \in \Delta_{+}}
\tilde{t}^{\pm}_{\pm\alpha}(\lambda)E_{\pm\alpha} ,  \qquad \tilde{S}^{\pm}(\lambda) =  \exp
\sum_{\alpha \in \Delta_{+}}\tilde{s}^{\pm}_{\pm\alpha}(\lambda)E_{\pm\alpha} 
 \end{array}
\end{equation}
where $\tilde{s}_{\pm\alpha}={s}_{\pm\alpha}$ and 
\begin{eqnarray}
&& \tilde{d}_1^{\pm}= {d}_1^{\pm}+\rmi\Phi, \quad  \tilde{d}_2^{\pm}= {d}_2^{\pm} -\rmi\Phi,
\qquad \tilde{t}_{\pm \alpha}^{\pm}= {t}_{\pm \alpha}^{\pm}\exp{\pm\rmi\alpha(J')\Phi}.
\end{eqnarray}
So setting $\tilde{S}^{\pm} (\lambda)={S}^{\pm} (\lambda)$ and
$\tilde{s}^{\pm}_{\pm \alpha}={s}^{\pm}_{\pm \alpha}$ one could put (\ref{eq:asipmtilde}) in a more symmetric form
\begin{equation}\label{eq:asipmtildesf}
\begin{array}{lr}
\rme^{\rmi\lambda J_0x}g_-^{-1} \tilde{\chi}^+(x,\lambda)\to \tilde{S}^{+} (\lambda),\qquad &x\to-\infty,\\[4pt]
\rme^{\rmi\lambda J_0x} g_+^{-1}\tilde{\chi}^+(x,\lambda)\to \tilde{T}^{-}(\lambda) \tilde{D}^{+}(\lambda),\qquad &x\to
+\infty,\\[4pt]
\rme^{\rmi\lambda J_0x} g_-^{-1}\tilde{\chi}^-(x,\lambda)\to \tilde{S}^{-} (\lambda),\qquad  &x\to-\infty,\\[4pt]
\rme^{\rmi\lambda J_0x}g_+^{-1}{\tilde\chi}^-(x,\lambda) \to \tilde{T}^{+} (\lambda)\tilde{D}^{-}(\lambda),\quad &x\to+\infty.
\end{array}
\end{equation}
One can see that again for $\alpha\in \Delta_{+}$
\begin{equation}\label{eq:stdsymmet}
\begin{array}{l}
\tilde{s}^{-}_{-\alpha}(-\lambda)=\tilde{s}^{+}_{-{\cal K}\alpha}(\lambda),\qquad  \tilde{t}^-_{{-\alpha}}(-\lambda)=\tilde{t}^{+}_{-{\cal K}\alpha}(\lambda),\\[4pt]
\tilde{s}^+_{{\alpha}}(-\lambda)=\tilde{s}^{-}_{{\cal K}\alpha}(\lambda),\qquad  \tilde{t}^+_{{\alpha}}(-\lambda)=\tilde{t}^{-}_{{\cal K}\alpha}(\lambda),\\[4pt]
\tilde{d}_{1}^+(-\lambda)=-\tilde{d}_2^{-}(\lambda),\qquad \tilde{d}_{2}^+(-\lambda)=-\tilde{d}_1^{-}(\lambda)
\end{array}
\end{equation}
and we get that the families
\begin{equation}
\begin{array}{l}
\tilde{{\cal F}}_s=\{\tilde{s}_{\alpha}^+(\lambda), \tilde{s}_{-\alpha}^{-}(\lambda),\quad  \lambda\in \bbr,\quad \alpha\in\Delta_{+}\},\\[6pt]
\tilde{{\cal F}}_t=\{\tilde{t}_{\alpha}^+(\lambda), \tilde{t}_{-\alpha}^{-}(\lambda),\quad  \lambda\in \bbr, \quad\alpha\in\Delta_{+}\}
\end{array}
\end{equation}
could be considered as scattering data for the $\rm GMV_{\pm}$ system. Taking into account that $\Phi$
is integral of motion for the NLEEs associated to the $\rm GMV_{\pm}$ system (see Remark \ref{Rem:intmot})
we notice that passing from quantities without tilde to quantities with tilde is essentially the same thing,
that is $S$ could be reconstructed using the same scattering data as the scattering data for $q$, something
that of course could be expected.

\subsection{Scattering Data Related to the Expansions over the Adjoint Solutions}

Now let us turn our attention to the scattering data associated with the expansions over adjoint solutions of 
the $\rm GMV_{\pm}$.  First we make the following observation. Suppose $B =\const\in \mathfrak{h}$. Then the
function $\tilde{B}=gBg^{-1}$ obviously satisfies $[S,\tilde{B}]=0$, so $\tilde{B}$ takes values in $\mathfrak{h}_S$.
Moreover, we have that $\langle\tilde{B},S\rangle=\langle B,J_0\rangle=\const$ and
$\langle\tilde{B},S_1\rangle=\langle B,J_1\rangle=\const$ where $J_1=J_0^2-\frac{2}{3}\mathbf{1}$. But $J_0$
and $J_1$ span $\mathfrak{h}$, so $B$ is of the form $B=a_0J_0+a_1J_1$ where $a_0, a_1$ are some numbers.
Then $\tilde{B}=a_0S+a_1S_1$. From the other side, we have seen that $S_x$ and $S_{1,x}$ are orthogonal to
$\mathfrak{h}_S$, so we see that $\tilde{B}_x$ is also orthogonal to   $\mathfrak{h}_S$, that is, belongs to
$\mathfrak{h}_S^{\perp}$. Take now any fundamental solution $\tilde{\chi}$ and a any root vector $E_{\alpha}$.
We have
\[
\langle \tilde{B}_x,\tilde{\chi} {E}_{\alpha}\tilde{\chi}^{-1}\rangle=\langle  \tilde{B}_x,\pi_S(\tilde{\chi} E_{\alpha}\tilde{\chi}^{-1})\rangle=\langle [S,\ad_S^{-1} \tilde{B}_x],\pi_S(\tilde{\chi} E_{\alpha}\tilde{\chi}^{-1})\rangle.\]
From the other side, since $\rmi\partial_x(\tilde{\chi}E_{\alpha}\tilde{\chi}^{-1})=\lambda[S,\tilde{\chi}E_{\alpha}\tilde{\chi}^{-1}]$,
we have that 
\[
\partial_x(\langle \tilde{B},\tilde{\chi} E_{\alpha}\tilde{\chi}^{-1}\rangle)=\langle \tilde{B}_x,\tilde{\chi}E_{\alpha}\tilde{\chi}^{-1}\rangle.
\]
Combining these relations we obtain that
\begin{equation}\label{eq:Wr1}
\displaystyle\int\limits_{-\infty}^\infty \left. \langle  [S,\ad_S^{-1} \tilde{B}_x], \pi_S(\tilde{\chi}
E_\alpha {\tilde{\chi}^{-1}}) \rangle \rmd x = \langle   \tilde{B},
{\tilde{\chi}} E_\alpha \tilde{\chi}^{-1} \rangle
\right |_{-\infty}^\infty , \quad \lambda \in \bbr,\quad
\alpha \in \Delta
\end{equation}
provided the integral exists, that is the right hand exists. Putting in the above formulas
$\tilde{\chi}=\tilde{\chi}^{\pm}$ we see that we can calculate the coefficients in the expansions from
Theorem  \ref{th:comprel} for the functions $\ad_S^{-1} \tilde{B}_x$. We present the result of these
calculations as a theorem.
\begin{theorem} \label{th: exppotenS}
The following expansion formulas hold:
\begin{equation}\label{eq:exppot}
\begin{array}{lc}
{\rm a)}&-\ad_S^{-1} \tilde{B}_x = \displaystyle \frac{1}{2 \pi} \displaystyle \int\limits_{-\infty}^\infty
 \left[ \sum\limits_{\alpha \in \Delta_{+}} \left(
\tilde{\rho}^{+}_{B,-\alpha} \tilde{\bf e}^{+}_\alpha -
\tilde{\rho}^{-}_{B,\alpha} \tilde{\bf e}^{-}_{-\alpha} \right) \right] \rmd
\lambda  \\[8pt]
{\rm b)}&-\ad_S^{-1} \tilde{B}_x = \displaystyle \frac{1}{2 \pi} \displaystyle \int\limits_{-\infty}^\infty
\left[ \sum\limits_{\alpha \in \Delta_{+}} \left(
\tilde{\sigma}^{+}_{B,\alpha} \tilde{\bf e}^{+}_{-\alpha} -
\tilde{\sigma}^{-}_{B,-\alpha} \tilde{\bf e}^{-}_{\alpha} \right) \right] \rmd
\lambda.
\end{array}
\end{equation}
The coefficients $\tilde{\rho}$ and $\tilde{\sigma}$ (of course we have $\alpha \in \Delta_{+}$ and $\lambda \in \bbr$)
are equal to 
\begin{equation}\label{eq:rho-sig}
\begin{array}{l}
\tilde{\rho}^\pm_{B,\mp\alpha} = \langle  (\tilde{S}^{\pm})^{-1}B
\tilde{S}^\pm , E_{\mp\alpha} \rangle = \langle  ({S}^{\pm})^{-1}B
{S}^\pm , E_{\mp\alpha} \rangle,\\ [8pt]
\tilde{\sigma}^\pm_{B,\pm\alpha} = \langle  (\tilde{D}^\pm)^{-1}
(\tilde{T}^\pm)^{-1} B~ \tilde{T}^\pm\tilde{D}^\pm , E_{\pm\alpha}
\rangle=\\[8pt]
\langle  ({D}^\pm)^{-1}({T}^\pm)^{-1} B~ {T}^\pm{D}^\pm , E_{\pm\alpha}\rangle . 
\end{array}
\end{equation}
\end{theorem}
The functions
\begin{equation}\label{eq:rho-sigcan}
\begin{array}{l}
{\rho}^\pm_{B,\mp\alpha}=\langle  ({S}^{\pm})^{-1}B
{S}^\pm , E_{\mp\alpha} \rangle,\\ [8pt]
{\sigma}^\pm_{B,\pm\alpha} =\langle  ({D}^\pm)^{-1}({T}^\pm)^{-1} B~ {T}^\pm{D}^\pm , E_{\pm\alpha}\rangle  
\end{array}
\end{equation}
are exactly the coefficients one obtains  expanding ${\rm i\ad}_{J_0}^{-1} [B,q]$ over the adjoint solutions
of (\ref{eq:GZSours}), see for example \cite{G86,Yan93}.  Thus we have 
\begin{equation}
\tilde{\rho}^\pm_{B,\mp\alpha} = {\rho}^\pm_{B,\mp\alpha},\qquad \tilde{\sigma}^\pm_{B,\pm\alpha} = {\sigma}^\pm_{B,\pm\alpha} .
\end{equation}
It is known that provided that $B$ is regular element, the families 
\begin{equation}\label{eq:scatdat-r-scan}
\begin{array}{lc}
{{\cal F}}_{\rho}=\{ \rho^{+}_{B,-\alpha}(\lambda), \quad
\rho^{-}_{B,\alpha}(\lambda):  \quad \lambda \in \bbr, \quad \alpha \in
\Delta_{+}\}, \\ [8pt] 
{{\cal F}}_{\sigma}= \{ \sigma^{+}_{B,\alpha}(\lambda),\quad
\sigma^{-}_{B,-\alpha}(\lambda): \quad \lambda \in \bbr, \quad\alpha \in
\Delta_{+} \} 
\end{array}
\end{equation} 
could also be taken as sets of scattering data for GZS (\ref{eq:GZSours}), that is using it one can
construct the potential $q$. Then the families
\begin{equation}\label{eq:scatdat-r-spole}
\begin{array}{lc}
\tilde{{\cal F}}_{\rho}=\{ \tilde{\rho}^{+}_{B,-\alpha}(\lambda),~
\tilde{\rho}^{-}_{B,\alpha}(\lambda):  \quad \lambda \in \bbr, \quad\alpha \in
\Delta_{+}\} ,\\ [8pt] 
\tilde{{\cal F}}_{\sigma}= \{ \tilde{\sigma}^{+}_{B,\alpha}(\lambda),~
\tilde{\sigma}^{-}_{B,-\alpha}(\lambda): \quad \lambda \in \bbr, \quad\alpha \in
\Delta_{+}\}
\end{array}
\end{equation}
could be treated as scattering data for the $\rm GMV_{\pm}$ system.  In particular, if one chooses $B=J_0$
and $B=J_1$ one will obtain expansions for the functions $\ad_S^{-1} {S}_x$ and $\ad_S^{-1} S_{1x}$. However,
depending on the properties of $B$ the scattering data have different symmetries. Indeed, suppose that
$h(\tilde{B})=\eta \tilde{B}$, where $\eta=\pm 1$. This as easily seen is equivalent to $k(B)=\eta B$.
Then of course we shall have also $h(\tilde{B}_x)=\eta \tilde{B}_x$ and
\begin{eqnarray*}
\langle\langle  [S,\ad_S^{-1} \tilde{B}_x], \pi_S\tilde{\chi}^{\pm}(x,\lambda)
E_\alpha ({\tilde{\chi}^{\pm})^{-1}}(x,\lambda) \rangle\rangle =\\
\eta\langle\langle  [S,\ad_S^{-1} \tilde{B}_x], \pi_S\tilde{\chi}^{\mp}
E_{{\mathcal K}\alpha} {(\tilde{\chi}^{\mp})^{-1}}(x,-\lambda) \rangle\rangle.
\end{eqnarray*}
Consequently, for $\alpha\in \Delta_{+}$ we have
\begin{equation}
\begin{array}{l}
\tilde{\rho}^{+}_{B,-\alpha}(\lambda)= \eta\tilde{\rho}^{-}_{B,-{\mathcal K}\alpha}(-\lambda), \qquad 
\tilde{\rho}^{-}_{B,\alpha}(\lambda)=\eta \tilde{\rho}^{+}_{B,{\mathcal K}\alpha}(\lambda),\\[6pt]
\tilde{\sigma}^{+}_{B,\alpha}(\lambda)= \eta\tilde{\sigma}^{-}_{B,{\mathcal K}\alpha}(-\lambda), \qquad
\tilde{\sigma}^{-}_{B,-\alpha}(\lambda)=\eta \tilde{\sigma}^{+}_{B,-{\mathcal K}\alpha}(-\lambda).
 \end{array}
\end{equation}
Since $k(J_0)=-J_0$ and $k(J_1)=J_1$, taking for example $B=J_0$, one obtains 
\begin{equation}
\begin{array}{l}
\tilde{\rho}^{+}_{J_0,-\alpha}(\lambda)= -\tilde{\rho}^{-}_{J_0,-{\mathcal K}\alpha}(-\lambda), \qquad 
\tilde{\rho}^{-}_{J_0,\alpha}(\lambda)=-\tilde{\rho}^{+}_{J_0,{\mathcal K}\alpha}(\lambda),\\[6pt]
\tilde{\sigma}^{+}_{B,\alpha}(\lambda)= -\tilde{\sigma}^{-}_{J_0,{\mathcal K}\alpha}(-\lambda), \qquad \tilde{\sigma}^{-}_{J_0,-\alpha}(\lambda)=-\tilde{\sigma}^{+}_{J_0,-{\mathcal K}\alpha}(-\lambda)
 \end{array}
\end{equation}
and one sees that
\begin{equation}\label{eq:exppotS}
\begin{array}{lc}
{\rm a)}&-\ad_S^{-1} \tilde{S}_x = \displaystyle \frac{1}{2 \pi} \displaystyle \int\limits_{-\infty}^\infty
 \left[ \sum\limits_{\alpha \in \Delta_{+}} \left(
\tilde{\rho}^{+}_{J_0,-\alpha} \tilde{\bf s}^{+}_\alpha -
\tilde{\rho}^{-}_{J_0,\alpha} \tilde{\bf s}^{-}_{-\alpha} \right) \right] \rmd
\lambda ,\\[8pt]
{\rm b)}&-\ad_S^{-1} \tilde{S}_x = \displaystyle \frac{1}{2 \pi} \displaystyle \int\limits_{-\infty}^\infty
\left[ \sum\limits_{\alpha \in \Delta_{+}} \left(
\tilde{\sigma}^{+}_{J_0,\alpha} \tilde{\bf s}^{+}_{-\alpha} -
\tilde{\sigma}^{-}_{J_0,-\alpha} \tilde{\bf s}^{-}_{\alpha} \right) \right] \rmd
\lambda
\end{array}
\end{equation}
as it should be according to the theory we developed. If we expand $\ad_S^{-1}S_{1x}$ then the expansion
could be reformulated as expansion over the functions $\tilde{\bf a}^{\pm}_{\alpha}$.

\section{Conclusions}

In the present work we have been able to do the following:
\begin{itemize}
\item We have showed that the results for the $\rm GMV=GMV_+$ system about the expansions over the adjoint
solutions could be generalized in two directions:
i) To apply for the system $\rm GMV_{-}$ corresponding to a pseudo-Hermitian reduction of a certain
$\asl(3, \mathbb{C})$-GZS system in pole gauge subject to $\mathbb{Z}_2\times \mathbb{Z}_2$ reduction of Mikhailov type. 
ii)  Both for $\rm GMV_{+}$  and $\rm GMV_{-}$ we have been able to develop the theory of expansions for
arbitrary constant asymptotic conditions of the potentials.
\item We have showed how the expansions  over adjoint solutions should be modified if one takes into account
possible symmetries of the functions we expand.  
\end{itemize}
The above was achieved using the gauge equivalence between the GZS system in pole gauge and the one in canonical
gauge, see Theorem \ref{th:GEq}. This has permitted to avoid repetition of cumbersome proofs and frequently to
reduce the things to purely algebraic arguments.

Since the expansions over the adjoint solutions of an auxiliary linear problem are the main tool of the so-called
recursion operator method to the soliton equations, it would be natural to develop the theory of these equations,
their conservation laws, hierarchies of compatible Poisson structures etc., for the GZS system
in pole gauge and in canonical gauge, making all the approach gauge-covariant. As mentioned in the introduction
this task that has been successfully achieved in the case of the Heisenberg Ferromagnet hierarchy and nonlinear
Schr\"odinger hierarchy and later generalized for systems on arbitrary simple Lie algebra with no reductions. 

Also, one must consider the discrete spectrum which we have not done here for the sake of brevity. 

Another direction in which we would like to proceed is construct special exact solutions for the first
equations in the hierarchies since they are likely to have some physical applications.

Including all these issues in one article is, of course, not possible, so we intend to address them in
future works.

\section*{Acknowledgments}
The work has been supported by the NRF incentive grant of South Africa and Grant DN 02-5 of Bulgarian Fund ``Scientific Research".

\end{document}